\newcommand{\beeq}{\begin{equation}}   
\newcommand{\eneq}{\end{equation}}
\newcommand{\be}{\begin{eqnarray}}
\newcommand{\ee}{\end{eqnarray}}
\newcommand{\bpic}{\begin{picture}}
\newcommand{\epic}{\end{picture}}
\newcommand{\bs}{\begin{scriptsize}}
\newcommand{\es}{\end{scriptsize}}

\def\dd{\partial} 
\def\la{\raise.16ex\hbox{$\langle$} \, }
\def\ra{\, \raise.16ex\hbox{$\rangle$} }

\documentstyle[preprint,aps,eqsecnum,epsfig,floats]{revtex}
\begin{document}
\draft 
\preprint{UFIFT-HET-02-35,~~NSF-ITP-02-176,~~LPT 03/05}
\vspace{2.5in}
\date{\today}
\tightenlines
\vskip 2cm

\title{Gravitational Lensing by Dark Matter Caustics}

\author{C. Charmousis$^a$, V. Onemli$^b$, Z. Qiu$^b$ and
P. Sikivie$^{b,c}$}
\address{$^a$LPT{\footnote{Unit\'e mixte du CNRS (UMR 8627).}},
B\^at. 210, Universit\'e de Paris-Sud, F-91405 Orsay, France\\
$^b$Department of Physics, University of Florida, 
Gainesville, FL 32611-8440\\
$^c$KITP, University of California, Santa Barbara, CA 93106-4030}

\maketitle

\begin{abstract}
Dark matter caustics have specific density profiles and, therefore,  
precisely calculable gravitational lensing properties.  We present a
formalism which simplifies the relevant calculations, and apply it to 
four specific cases.  In the first three, the line of sight is tangent 
to a smooth caustic surface.  The curvature of the surface at the tangent 
point is positive, negative or zero.  In the fourth case the line of 
sight passes near a cusp.  For each we derive the map between the image 
and source planes.  In some cases, a point source has multiple images 
and experiences infinite magnification when the images merge.  
Unfortunately, for the dark matter caustics expected in realistic 
galactic halo models, the angular resolution required to resolve 
the multiple images is not presently achievable.  A more promising 
approach aims to observe the distortions caused by dark matter 
caustics in the images of extended sources such as radio jets.
\end{abstract}

\narrowtext

\section{Introduction}
\label{sec:in}

Gravitational lensing techniques have proven useful in studying the
distribution of dark matter in the universe.  They have been used to
reveal the existence of massive compact halo objects (MACHOs) in galaxies
\cite{macho}, and to constrain the mass distribution in galaxy clusters
\cite{clu}.  In this paper, we calculate the gravitational lensing
properties of dark matter caustics \cite{cr,sing,Tre}.  Gravitational
lensing by dark matter caustics has been discussed by Hogan \cite{Hogan}.  
We confirm Hogan's results for the case he considered, called the
`concave fold' in our nomenclature.  We study additional cases and
introduce a formalism to facilitate the calculations.

In the cold dark matter (CDM) cosmology, caustics in the distribution 
of dark matter are plentiful once density perturbations enter the 
non-linear regime.  A dark matter caustic is a surface in physical 
space where the dark matter density is large because the sheet on 
which the particles lie in phase-space has a fold there.  Dark matter
caustics can move about but they are stable.

To see the validity of the preceding statements, note that before the
onset of galaxy formation the cold dark matter particles lie on a thin 
3-dim. sheet in 6-dim. phase-space.  The thickness of this sheet 
is the primordial velocity dispersion $\delta v$ of the particles. 
It is of order \cite{sing}
\begin{equation}
\delta v_a (t) \sim 3\cdot 10^{-17} \left( {10^{-5} {\rm eV} \over
m_a}\right)^{5/6}~\left({t_0\over t}\right)^{2/3}
\label{vda}
\end{equation}
for axions, and
\begin{equation}
\delta v_W (t) \sim 10^{-11} \left({{\rm GeV} \over m_W}\right)^{1/2}
\left({t_0\over t}\right)^{2/3}
\label{vdW}
\end{equation}
for weakly interacting massive particles (WIMPs), such as the 
lightest supersymmetric particle in supersymmetric extensions 
of the Standard Model.  In Eqs. (1.1-2), $t_0$ is the 
present age of the universe and $m_a$ and $m_W$ are respectively 
the masses of the axion and WIMP.  The present average axion number
density is
\begin{equation}
n_a(t_0) = {1.5~10^{64} \over {\rm pc}^3}\, ~\Omega_a~
\left({h \over 0.7}\right)^2~\left({10^{-5}{\rm eV} \over m_a}\right)
\label{na}
\end{equation}
where $\Omega_a$ is their present energy density in units of the 
critical density, and $h$ is the Hubble expansion rate in units of 
100 km/(s$\cdot$Mpc).  Likewise, for WIMPs
\begin{equation}
n_W(t_0) = {1.5~10^{50} \over {\rm pc}^3}\, ~\Omega_W~
\left({h \over 0.7}\right)^2~\left({{\rm GeV} \over m_W}\right)~~~\ .
\label{nW}
\end{equation}
The large exponents in Eqs. (1.3-4) indicate that the density of cold 
dark matter particles is enormous in terms of astronomical length scales.  
Being effectively collisionless, the particles experience only
gravitational forces. These are universal and vary only on huge distances
compared to the 
interparticle distance.  Hence the sheet on which the particles lie 
in phase-space is continuous.  It cannot break and therefore
its evolution is constrained by topology.

In contrast, in present numerical simulations of galactic halo formation, 
the particle mass $m$ is of order $10^6~M_\odot$ and hence the 
number density
\begin{equation}
n(t_0) =  {1.4~10^{-13} \over {\rm pc}^3}\, ~\Omega
~\left({h \over 0.7}\right)^2~\left({10^6~M_\odot \over m}\right)~~~\ .
\label{nsim}
\end{equation}
The resolution of the simulations is inadequate to describe the 
dynamical evolution of the phase-space sheet and the physical 
implications of its existence.

Where a galaxy forms, the sheet wraps up in phase-space,
turning clockwise in any two dim. cut $(x, \dot{x})$ of that   
space.  $x$ is the physical space coordinate in an arbitrary direction and
$\dot{x}$ its associated velocity.  The outcome of this process is a
discrete set of flows at any physical point in a galactic halo \cite{ips}.
Two flows are associated with particles falling through the galaxy for
the first time ($n=1$), two other flows are associated with particles
falling through the galaxy for the second time ($n=2$), and so
on.  Scattering in the gravitational wells of inhomogeneities in  
the galaxy (e.g. molecular clouds and globular clusters) are
ineffective in thermalizing the flows with low values of $n$.

Caustics appear wherever the projection of the phase-space sheet onto
physical space has a fold \cite{cr,sing,Tre}.  Generically, caustics
are surfaces in physical space.  They separate regions with differing
number of flows.  On one side of a caustic surface there are two more
flows than on the other.  Because caustic surfaces occur wherever
the number of flows changes, they are topologically stable, in direct
analogy with domain walls. At the caustic, the dark matter density is very
large.  

There are two types of caustics in the halos of galaxies, inner and outer.  
The outer caustics are simple fold ($A_2$)  catastrophes located on
topological spheres surrounding the galaxy.  They occur where a given
outflow reaches its furthest distance from the galactic center before
falling back in.  The inner caustics are rings \cite{cr}.  They are
located near where the particles with the most angular momentum in a given
inflow reach their distance of closest approach to the galactic center
before going back out.  A caustic ring is a closed tube whose
cross-section is a $D_{-4}$ (also called {\it elliptic umbilic})
catastrophe \cite{sing}.  The existence of these caustics and their
topological properties are independent of any assumptions of symmetry.

Dark matter caustics have very well defined density profiles, and hence 
calculable gravitational lensing signatures \cite{Hogan}.  It is the
purpose of this paper to derive these signatures in a number of specific 
cases.  In the limit of zero velocity dispersion ($\delta v = 0$), the
density diverges when one approaches a caustic surface, on the side which 
has two extra flows, as the inverse square root of the distance to the
surface. This divergence is cut off if there is velocity dispersion, 
because the location of the caustic surface gets smeared over some
distance $\delta x$.  For the dark matter caustics in galactic halos, 
$\delta x$ and $\delta v$ are related by \cite{cr}
\begin{equation}
\delta x \sim {R~\delta v \over v}
\label{sme}
\end{equation}
where $v$ is the order of magnitude of the velocity of the particles 
in the flow and $R$ is the distance scale over which that flow turns
around, i.e. changes its direction.  For a galaxy like our own, 
$v = 500$ km/s and $R = 200$ kpc are typical orders of magnitude.   

As was mentioned earlier, the primordial velocity dispersion of the   
leading cold dark matter candidates is very small.  Using 
Eqs. (1.1-2), one finds that axion caustics in galactic 
halos are typically smeared over 
\begin{equation}
\delta x_a \sim 10^{10}\, {\rm cm} 
\left({10^{-5} {\rm eV} \over m_a} \right)^{5/6}
\label{dxa}
\end{equation}
as a result of their primordial velocity dispersion, whereas WIMP 
caustics are smeared over 
\begin{equation}
\delta x_W \sim 3 \cdot 10^{15}\, {\rm cm} 
\left({{\rm GeV} \over m_W}\right)^{1 \over 2}~~~~\ .
\label{dxW}
\end{equation}
It should be kept in mind, however, that a cold dark matter flow may have
an effective 
velocity dispersion which is larger than its primordial velocity 
dispersion.  Effective velocity dispersion occurs when the sheet 
on which the dark matter particles lie in phase-space is wrapped up on 
scales which are small compared to the galaxy as a whole.  It is 
associated with the clumpiness of the dark matter falling onto the
galaxy.  The effective velocity dispersion of a 
flow may vary from point to point, taking larger values where more small
scale structure has formed, and taking the minimum primordial value where 
no small scale structure has formed.  For a coarse-grained observer, the 
dark matter caustic is smeared over $\delta x$ given by Eq. (\ref{sme}) 
where $\delta v$ is the effective velocity dispersion of the flow.  

Little is known about the size of the effective velocity dispersion of
dark matter flows in galactic halos.  However, in ref. \cite{milk}, a
triangular feature in the IRAS map of the Milky Way was interpreted as the
imprint upon baryonic matter of the caustic ring of dark matter nearest to
us.  The sharpness of the feature's edges implies an upper limit of 20 pc
on the distance $\delta x$ over which that caustic is smeared, and hence
an upper limit of order 50 m/s on the effective velocity dispersion of the
corresponding flow.

The gravitational lensing effects of a caustic surface are largest when
the line of sight is near tangent to the surface because the contrast in
column density is largest there.  The effects depend on the curvature of
the caustic surface at the tangent point in the direction of the line of
sight: the smaller the curvature, the larger the effects.  A caustic is an
oriented surface because one side has two more flows than the other.  We
will consider three cases of gravitational lensing by a smooth caustic
surface.  In the first case, the line of sight is near tangent to a
caustic surface which curves towards the side with two extra flows; see
Fig. 4.  We call such a surface `concave'. In the second case, the surface
is `convex', i.e. it curves away from the side with two extra flows; see
Fig. 6.  In the third case, the caustic surface has zero curvature at the
tangent point (the radius of curvature is infinite), but the tangent line
is entirely outside the side with two extra flows.

Caustic surfaces may have cusps.  The outer dark matter caustics of 
galactic halos are topological spheres, which have no cusps.  But 
the inner dark matter caustics of galactic halos are closed tubes 
whose cross-section is a $D_{-4}$ catastrophe; see Fig. 2.  This
cross-section has three cusps.  We call it a `tricusp'. The fourth 
case of gravitational lensing which we consider has line of sight 
near a cusp, and parallel to the plane of the cusp; see Fig. 9.

Gravitational lensing produces a map of an object surface onto an 
image surface.  The magnification is the inverse of the Jacobian 
of this map.  Because dark matter caustics have well defined 
density profiles, it is a neat mathematical exercise to calculate 
their gravitational lensing characteristics.  The images of 
extended sources may show distortions that can be unambiguously
attributed to lensing by dark matter caustics in the limit of 
perfect observations.  We will see that in three of the cases 
considered, a point source can have multiple images.  In those 
cases, when two images merge, the Jacobian of the map vanishes 
and the magnification diverges.  So, at least in theory, it seems 
that gravitational lensing is a very attractive tool for investigating 
dark matter caustics.  Observation of the calculated lensing 
signatures would give direct evidence for caustics and CDM.

We have been particularly motivated by the possibility \cite{Hogan}
that the observer might be able to distinguish between axions and WIMPs 
by determining the distance over which the caustics are smeared.  The
nearby caustic, whose position is revealed according to ref. \cite{milk}
by a triangular feature in the IRAS map of the Milky Way plane, is only 1
kpc away from us in the direction of observation.  By observing the
gravitational lensing due to that caustic, one may be able to measure
$\delta x$ as small as $10^{13}$ cm, assuming an angular resolution of $3
\cdot 10^{-9}$ radians.  If $\delta x$ turned out to be that small, the
WIMP dark matter hypothesis would be severely challenged; see Eq.
(\ref{dxW}).  Unfortunately, as will be shown below, the gravitational
lensing due to a caustic only a kpc away from us is too weak to be
observed with present instruments.  It is well known that gravitational
lensing effects are proportional to ${D_L D_{LS} \over D_S}$ where $D_S,
D_L$ and $D_{LS}$ are respectively the distances from the observer to the
source, from the observer to the lens and from the lens to the source.  
We will see below that, for the gravitational lensing effects of dark
matter caustics to be observable with present technology, the lenses and
sources must be as far as possible, at the cosmological distances of order
Gpc.  Even then, the observation of such effects will be difficult.  
Unfortunately, at Gpc distances it is not possible to measure $\delta x$
as small as Eqs. (1.7-8) with foreseeable technology.  So it seems
unlikely that one will be able to distinguish between dark matter
candidates on the basis of the gravitational lensing characteristics of
the caustics they form.  Henceforth, unless otherwise stated, the velocity
dispersion is set equal to zero. 

The remainder of this paper is organized as follows.  In section II, we
describe the outer and inner caustics of dark matter in galactic halos.  
We provide estimates of their sizes and of the strengths of their density
profiles, using the self-similar infall model of galactic halo formation.  
In section III, the general formalism of gravitational lensing is
reviewed.  We show how the calculations can be streamlined for the case of
gravitational lensing by dark matter caustics.  In section IV, we
calculate the gravitational lensing properties of dark matter caustics in
the four cases mentioned above.  In section V, we summarize our
conclusions.

\section{Density profiles of caustics}
\label{sec:dmc}

In this section we discuss the density profiles of the dark matter 
caustics present in galactic halos.  We consider the outer caustics 
first.  

\subsection{Outer caustics}

Outer caustics are closed surfaces (topological spheres) near the
$(n+1)$th turn-around radii with $n=1, 2, 3, ...\;$. Indeed the number of
flows changes by two there because of the fall back of the particles. There 
is no caustic associated with the first turn-around, see Fig. 1. The outer
caustics are described by simple ``fold'' ($A_2$) catastrophes. Their
density profile is:
\be
d_n (\sigma)=\frac{A_n}{\sqrt{\sigma}}\,\Theta (\sigma)
\label{densityprofile}
\ee
for small $\sigma$, where $\sigma$ is the distance to the caustic, $\Theta
(\sigma)$ is the Heaviside function, and $A_n$ is a constant which we 
call the {\it fold coefficient}. $\sigma >0$ on the side with two extra 
flows, i.e. towards the galactic center. Therefore, when an outer caustic 
is approached from the inside the density diverges as
$\sim\frac{1}{\sqrt{\sigma}}$, abruptly falling to zero on the outside. 
The observation \cite{malin} of arc-like shells surrounding giant 
elliptical galaxies can be interpreted \cite{quinn} as outer caustics 
in the distribution of baryonic matter falling onto those galaxies.

To estimate $A_n$ in Eq. (\ref{densityprofile}), consider the 
time evolution of CDM particles which are falling out of a galactic
halo for the $n$th time and then fall back in.  Let $R_n$ be the
turn-around radius. We assume that the rotation curve of the galaxy 
is flat near $r=R_n$ with time-independent rotation velocity $v_{\rm
rot}$.  The
gravitational potential is then:
\be
V(r)=v^2_{\rm\tiny{rot}}\ln\left(\frac{r}{R_n}\right) ~~~\ .
\ee
The particles have trajectory $r(t)$ such that
\be
\left|\frac{dr}{dt}\right|=\sqrt{2\left(E-V(r)\right)}=   
v_{\rm\tiny{rot}}\sqrt{2\ln\left(\frac{R_n}{r}\right)} ~~~\ ,
\label{drdt}
\ee
where $E$ is the energy per unit mass.  Eq. (\ref{drdt}) neglects 
the angular momentum of the particles.  This is a good approximation at
turnaround since the particles are far from their distance of closest 
approach to the galactic center.  Finally, we assume that the flow is 
stationary.  In that case the number of particles flowing per unit solid 
angle and per unit time, $\frac{dN}{d\Omega dt}$, is independent of $t$ 
and $r$, and the caustic is located exactly at the $(n+1)$th turn-around 
radius $R_n$.

Let us emphasize that none of the assumptions - spherical symmetry, flat
rotation curve, time independence of the gravitational potential,
radial orbits, and stationarity of the flow - affect the existence of 
the outer caustics or the fact that they have the density profile given
by Eq. (\ref{densityprofile}). The assumptions are made only to obtain
estimates of the coefficients $A_n$. 

The mass density of particles, $d_n(r)$, follows from the equality:
$2\frac{dN}{d\Omega dt}dt= {d_n(r) \over m} r^2 dr$, where $m$ is the
mass of each particle.  The factor of $2$ appears because there are two
distinct flows, out and in. Using Eq. (\ref{drdt}) we obtain the
density distribution near the $n$th outer caustic:
\be
d_n(r)=2\frac{dM}{d\Omega
dt}\Bigg|_{n}\,\,\frac{1}{r^2
v_{\rm\tiny{rot}}\sqrt{2\ln{\left(\frac{R_n}{r}\right)}}} ~~~\ ,
\label{dnr}
\ee
where $\frac{dM}{d\Omega dt}\equiv m\frac{dN}{d\Omega dt}$. Near the caustic, 
$\ln{\left(\frac{R_n}{r}\right)}=\frac{\sigma}{R_n}$ where 
$\sigma = R_n - r$.  Inserting this into Eq. (\ref{dnr}) and 
comparing with Eq. (\ref{densityprofile}), we find:
\be
A_n=\frac{\sqrt{2}}{v_{\rm\tiny{rot}}}\frac{dM}{d\Omega
dt}\Bigg|_n R^{-3/2}_n ~~~\ .  
\label{AN}
\ee 
Estimates of $R_n$ and $\frac{dM}{d\Omega dt}\Big|_n$ can be extracted
from ref. \cite{STW} for the case of self-similar infall \cite{FG,B}.
The infall is called {\it self-similar} if it is time-independent
after all distances are rescaled by the turn-around radius $R(t)$ at time
$t$ and all masses are rescaled by the mass $M(t)$ interior to $R(t)$. In
the case of zero angular momentum and spherical symmetry, the infall is
self-similar if the initial overdensity profile has the form 
$\frac{\delta M_i}{M_i}=\left(\frac{M_0}{M_i}\right)^\epsilon$ where 
$M_0$ and $\epsilon$ are parameters \cite{FG}. In CDM theories of large 
scale structure formation, $\epsilon$ is expected to be in the range $0.2$
to $0.35$ \cite{STW}. In that range, the galactic rotation curves
predicted by the self-similar infall model are flat \cite{FG}.

The $R_n$ and $\frac{dM}{d\Omega dt}\Big|_n$ do not depend sharply 
upon $\epsilon$.  Our estimates are for $\epsilon = 0.2$ because they 
can be most readily obtained from ref. \cite{STW} in that case.
For $\epsilon=0.2$, the radii of the $(n+1)$th turn-around spheres 
are approximately
\be
\{R_n: n=1,2,...\} \simeq (240,~120,~90,~70,~60,...)~{\rm kpc}
\cdot\left(\frac{v_{\rm{\tiny rot}}}{220\,{\tiny{\rm km/s}}}\right)
\left(\frac{0.7}{h}\right)~~~\ .
\label{Rn}
\ee
Moreover, one has
\be
\left.\frac{dM}{d\Omega dt} \right|_n\frac{1}{v_{\rm{\tiny rot}}}
=F_n\,\sqrt{2}\,\frac{v^2_{\rm {\tiny rot}}}{4\pi G}~~~\ ,
\label{Fn}
\ee
with 
\be
\{F_n : n=1,2,...\} \simeq (20,~8,~5,~4,~3, ...)~10^{-2}~~~\ .
\label{Fne}
\ee
Combining Eqs. (\ref{AN} - \ref{Fne}), we find
\be
\{A_n :n=1,2,..\} \sim (2,~2,~2,~3,~3, ...)\cdot 
{\rm \frac{10^{-5} \,gr}{cm^2\,kpc^{1/2}}}\nonumber\\
\cdot\left(\frac{v_{\rm{\tiny rot}}}{220\,{\tiny{\rm km/s}}}\right)^{1/2}
\left(\frac{h}{0.7}\right)^{3/2}\; .
\label{An}
\ee 
Generally the outer caustics are 'concave', i.e. they are curved 
towards the side which has two extra flows.  Their radius of curvature 
is of order $R_n$.  The lensing by concave caustic 
surfaces is discussed in section IV.A.

In the next subsection, we discuss the density profiles of inner 
caustics.

\subsection{Inner caustics}

The inner caustics \cite{cr} are closed tubes whose cross-section, 
shown in Fig. 2, is a $D_{-4}$ catastrophe \cite{sing}. They are 
located near where the particles with the most angular momentum 
in a given inflow are at their distance of closest approach 
to the galactic center. We call them `caustic rings'.  For 
simplicity, we study caustic rings which are axially symmetric 
about the $z$-direction as well as reflection symmetric with 
respect to the $z=0$ plane.  The dark matter flow is then 
effectively 2-dimensional.  In galactocentric cylindrical
coordinates, the flow at such a caustic ring is described by
\cite{sing}:
\be
z(\alpha , \tau) =b\,\alpha\,\tau \hskip 1cm \mbox{and} \hskip 1cm
\rho(\alpha , \tau) =a+{u\over 2}(\tau -\tau_0)^2-{s\over 2}\alpha^2~~~\ ,
\label{flow}
\ee  
to lowest order in an expansion in powers of $\alpha$ and $\tau$.
$\alpha$ and $\tau$ are parameters labeling the particles in the flow. 
$\rho$ is distance to the $z$-axis.  $\tau$ is the time a particle 
crosses the $z=0$ plane.  $\alpha$ is the declination of the particle, 
relative to the $z=0$ plane, when it was at last turnaround.  $b,~a,~u,
~\tau_0$ and $s$ are constants characteristic of the caustic ring.  $b,
~u,~s$ and $a$ are positive, whereas $\tau_0$ can have either sign.  $a$
is the radius of the ring; see Fig. 2.  The transverse dimensions of the 
ring are $~p={1 \over 2} u \tau_0^2~$ and 
$~q = {\sqrt{27} \over 4} {b \over \sqrt{us}} p~$ in the $\hat{\rho}$ 
and $\hat{z}$ directions respectively.  The cross-section of the ring  
has three cusps.  So we call it the ``tricusp''. The cusps are topological 
properties of the caustic.  Their presence does not depend on any 
assumptions of symmetry. 

The physical space density is given by \cite{sing}
\be
d(\rho,z)=\frac{1}{\rho}\sum_{j=1}^{n}\frac{dM}{d\Omega d\tau}(\alpha\, ,
\tau){\left.\frac{\cos{(\alpha)}}{\left|D_2 (\alpha\, ,
\tau)\right|}\right|_{\left(\alpha_j(\rho,z)\, , \,
\tau_j(\rho, z)\right)}}~~~\ , 
\label{km}
\ee
where $\frac{dM}{d\Omega d\tau} = 
\frac{dM}{2 \pi \cos{(\alpha)} d\alpha d\tau}$ is the mass falling in 
per unit time and unit solid angle. $\alpha_j(\rho,z)$ and 
$\tau_j(\rho, z)$, with $j = 1, ..., n$, are the solutions of 
$\rho(\alpha, \tau) = \rho$ and $z(\alpha, \tau) = z$.  $n(\rho, z)$ 
is the number of flows at $(\rho,z)$.  $n=2$ outside the caustic tube, 
whereas $n=4$ inside. $D_2(\alpha,\tau)\equiv\det \cal{D}(\alpha,\tau)$
with 
\be
\cal{D}(\alpha,\tau)\equiv\left(\begin{array}{cc}
{\dd\rho\over\dd\alpha}&~~~{\dd z\over\dd\alpha}\\
\\
{\dd\rho\over\dd\tau}&~~~ {\dd z\over\dd\tau}\\
\end{array}
\right)=\left(\begin{array}{cc}
{-s\alpha}&~~~{b\tau}\\
\\
{u(\tau-\tau_0)}&~~~ {b\alpha}\\
\end{array}
\right)
\label{Jaco}
\ee
is the Jacobian of the map $(\alpha ,\tau)\rightarrow (\rho, z)$.

In the limit of zero velocity dispersion, the density of dark matter 
particles is infinite at the location of caustic surfaces. Thus, the
location of the caustic ring surface is obtained by demanding that 
\be
D_2(\alpha,\tau)=-b[u\tau(\tau - \tau_0)+s\alpha^2]= 0 ~~~\ .
\label{jaco}
\ee
This implies
\be
\alpha(\tau) = \pm \sqrt{{u \over s} \tau (\tau_0 - \tau)}
\label{aoft}
\ee
with $0 \leq {\tau \over \tau_0} \leq 1$, and hence \cite{sing}
\be
\rho(\tau)=a+{u\over 2}(\tau-\tau_0)(2\tau-\tau_0) ~\ ,\qquad 
z(\tau)= \pm b\sqrt{{u\over s}\tau^3(\tau_0-\tau)}~~~\ ,
\label{parametric}
\ee
for the location of the tricusp outline.  Near the surface of a caustic 
ring, but away from the cusps, the density profile is that of a simple
fold:  $d(\sigma)=\frac{A}{\sqrt{\sigma}}\Theta(\sigma)$, with $\sigma>0$
inside the tricusp.  In the next three subsections we calculate the
fold  coefficient $A$ and the principal curvature radii, $R_1$ and $R_2$,
at arbitrary points on the surface of a caustic ring, other than the cusps.  
As a warm-up, we start with a special point, marked by a star in
Fig. 2.  In the fourth subsection, we obtain the density profile near the 
cusps.

\subsubsection{A sample point}
\label{subsection:aspecialpoint}

As an example, we determine the fold coefficient $A$ at $(\rho ,
z)=(a, 0)$;  
see Fig. \ref{fig:tricusp}.  Setting $\alpha=z=0$, 
$\sigma=\rho-a=\frac{1}{2}u(\tau-\tau_0)^2$ and $\tau\simeq\tau_0$, 
we have
\be
|D_2(\tau)|\simeq 2 b\sqrt{p\sigma} ~~~\ .
\ee
Including the factor 2 for in and
out flows, Eq. (\ref{km}) yields:  
\be
d(\sigma)=\frac{dM}{d\Omega
dt}\frac{1}{a b}\frac{1}{\sqrt{p\sigma}} ~~~\ .
\ee
Therefore,
\be
A_0=\frac{d^2M}{d\Omega
dt}\frac{1}{a b}\frac{1}{\sqrt{p}}~~~\ , 
\label{A_nc}
\ee
where the ${\tiny 0}$ subscript is to indicate that $A$ is being 
evaluated at the sample point.

To obtain a numerical estimate for $A_0$, we again use the self-similar 
infall model \cite{FG,B,STW} with $\epsilon = 0.2$.  For the $n$th ring,
we have \cite{cr}
\be
\{a_n: n=1, 2,...\}\simeq(39,~19.5,~13,~10,~8,...){\rm kpc}\,\cdot
\left(\frac{j_{\rm\tiny{max}}}{0.27}\right)\left(\frac{0.7}{h}\right)
\left(\frac{v_{\rm\tiny{rot}}}{220\,{\rm km/s}}\right)\; ,
\label{a_n}
\ee
where $j_{\rm \tiny{max}}$ is a parameter, with a specific value for each
halo, which is proportional to the amount of angular momentum that the 
dark matter particles have \cite{STW}.  Also,
\be
\left.\frac{d^2 M}{d\Omega dt}\right|_n
=f_n v_n\frac{v^2_{\bs\mbox{rot}\es}}
{4\pi G}~~~\ ,
\label{f}
\ee
where $v_n$ is the velocity of the particles in the $n$th caustic
ring, and the dimensionless coefficients $f_n$ characterize the 
density of the $n$th in and out flow.  In the self-similar model
\cite{cr}
\be
\{f_n: n=1, 2, ...\}\simeq (13,~5.5,~3.5,~2.5,~2,...)\cdot 10^{-2}
\ee
for $\epsilon=0.2$.  The $f_n$ are like the $F_n$ in Eq. (\ref{Fn}), but 
they describe the $n$th in and out flow near the caustic ring whereas
the $F_n$ describe that flow near turnaround. 

Combining Eqs. (\ref{A_nc}) and (\ref{f}) we have
\be
A_{0,n} =\frac{v^2_{\bs \mbox{rot}\es}}{4\pi
G}\,\frac{f_n}{a_n}\,\frac{v_n}{b_n}\,\frac{1}{\sqrt{p_n}}~~~\ .
\label{Anv2}
\ee
It was shown in ref. \cite{sing} that $b_n$ and $v_n$ are of the 
same order of magnitude.  Also, in ref. \cite{milk}, ten rises in 
the rotation curve of the Milky Way were interpreted as the effect 
of caustic rings.  In that case, the widths
$p_n$ of caustic rings are determined from the observed widths of 
the rises.  Typically one finds $p_n\sim 0.1\, a_n$.  Using this 
and $v_n \sim b_n$, Eq. (\ref{Anv2}) yields the estimates 
\be
\{A_{0,n}: n=1, 2,...\}\sim (3,~4,~4,~5,~5,...)\cdot
\frac{10^{-4}\,{\rm gr}}{\rm cm^2\,kpc^{1/2}}\nonumber\\
\cdot\left(\frac{0.27}{j_{\rm\tiny{max}}}\right)^{3/2}
\left(\frac{h}{0.7}\right)^{3/2}
\left(\frac{v_{\bs\mbox{rot}\es}}{220\, {\rm km/s}}\right)^{1/2}~~~\ .
\label{Anring}
\ee
At the point under consideration, $(\rho , z)=(a, 0)$, the surface of the
caustic ring is convex for all lines  of sight, i.e. all tangents at that
point are on the side with two extra flows.  If the line of sight is in 
the $z=0$ plane, the curvature radius is $a$.  If the line of sight is
perpendicular to the $z=0$ plane, the curvature radius is 
$2 {b^2 \over su} p$. Lensing by a convex caustic surface is 
discussed in subsection IV.B.

To obtain the lensing properties of the caustic ring surface at an 
arbitrary point, we need the coefficient $A$ and the 
curvature radii $R_1$ and $R_2$ at all locations.  We derive these
quantities in the next two subsubsections.

\subsubsection{The fold coefficient everywhere}

We choose an arbitrary point on the tricusp, i.e. on the surface of 
the caustic ring.  Its parameters are $(\alpha_1, \tau_1)$ with 
$\alpha_1$ given in terms of $\tau_1$ by Eq. (\ref{aoft}).  The
physical coordinates $(\rho_1, z_1)$ are given in terms of $\tau_1$
by Eq. (\ref{parametric}). We assume that the point is not 
at one of the three cusps.  The latter are located at $\tau_1=0$, at 
$\tau_1=\frac{3}{4}\tau_0$ with $\alpha_1>0$, and at 
$\tau_1=\frac{3}{4}\tau_0$ with $\alpha_1<0$. 

The vanishing of
$D_2(\tau_1)$ implies the existence of a zero eigenvector of the matrix 
\be
{\cal D}(\tau_1)=\left(\begin{array}{cc}
{-s\alpha_1}&{b\tau_1}\\
{u(\tau_1-\tau_0)}& {b\alpha_1}\\
\end{array}
\right)~~~\ .
\ee
Let us define $\theta_1(\tau_1)$ such that
\be
{\cal D}(\tau_1)\left(\begin{array}{c}
{\sin{(\theta_1)}}\\
{\cos{(\theta_1)}}\\
\end{array}
\right)=0~~~\ .
\label{dt1}
\ee
We have
\be
\sin{(\theta_1)}=\frac{b\tau_1}{\sqrt{(b\tau_1)^2+(s\alpha_1)^2}}
\hskip 0.5 cm \mbox{and} \hskip 0.5 cm 
\cos{(\theta_1)}=\frac{s\alpha_1}{\sqrt{(b\tau_1)^2+(s\alpha_1)^2}}~~~\ .
\ee
We define new Cartesian coordinates $(\sigma , \eta)$ related to 
$(\rho - \rho_1, z - z_1)$ by a rotation of angle $\theta_1+{\pi\over 2}$:
\be
\left(\begin{array}{c}                      
{\sigma}\\
{\eta}\\
\end{array}
\right)=  \left(\begin{array}{cc}
{-\sin{(\theta_1)}}&{-\cos{(\theta_1)}}\\
{\cos{(\theta_1)}}&{-\sin{(\theta_1)}}\\
\end{array}
\right)\left(\begin{array}{c}
{\rho-\rho_1}\\
{z -z_1}\\
\end{array}
\right)\; .
\label{mateq}
\ee
We now show that $\sigma$ is the coordinate in the direction 
orthogonal to the caustic surface at $(\rho_1, z_1)$.

Consider small deviations about $(\alpha_1 , \tau_1)$ in parameter
space: $(\alpha,\tau)=(\alpha_1+\Delta \alpha, \tau_1+\Delta \tau)$. 
Eq. (\ref{Jaco}) implies,
\be
\left(\begin{array}{c}
{{\Delta\rho}}\\
{{\Delta z}}\\
\end{array}
\right)= {\cal D}^T(\tau_1)\left(\begin{array}{c}
{{\Delta\alpha}}\\
{{\Delta\tau}}\\
\end{array}
\right)+O(\Delta\alpha^2 , \Delta\tau^2 ,
\Delta\alpha\Delta\tau)~~~\ ,
\label{DrDz}
\ee
where $T$ indicates transposition. The expansion of $\sigma$ in 
powers of $\Delta \alpha$ and $\Delta \tau$ yields:
\be
\sigma=O(\Delta\alpha^2 , \Delta\tau^2 ,
\Delta\alpha\Delta\tau)\; ,
\ee   
because the first order terms vanish:
\be
\left.\frac{\dd \sigma}{\dd
\alpha}\right|_{(\alpha_1,\tau_1)}\Delta \alpha+\left.\frac{\dd \sigma}
{\dd \tau}\right|_{(\alpha_1,\tau_1)}\Delta \tau = -(\Delta\alpha\,\,\,\,
\Delta\tau)~{\cal D}(\tau_1)\left(\begin{array}{c}
{\sin{(\theta_1)}}\\
{\cos{(\theta_1)}}\\
\end{array}
\right)=0~~~\ .
\ee
The fact that $\sigma$ is second order in $\Delta\alpha$ and $\Delta\tau$ 
shows that $\sigma$ is the coordinate in the direction perpendicular to
the caustic surface,  and
\be
\hat{\sigma}=-\sin{\left(\theta_1\right)}\hat{\rho}
-\cos{\left(\theta_1\right)}\hat{z}
\label{s}
\ee
is the unit normal to the surface, pointing inward. $\theta(\tau_1)$ is
the angle between the $\rho$ axis and the tangent line at $(\rho_1, z_1)$. 
See Fig. 3.
 
To obtain the density profile of the caustic near the point under
consideration, we need $D_2(\eta, \sigma)$ to order $\sqrt{\sigma}$.
So we calculate $\sigma$ to second order in powers of $\Delta \alpha$ 
and $\Delta \tau$, and $D_2$ and $\eta$ to first order.  We find
\be
\sigma=\frac{1}{2}(\Delta\alpha\,\,\,\,\Delta\tau)
\left(\begin{array}{cc}
{s\sin{(\theta_1)}}&{-b\cos{(\theta_1)}}\\ 
{-b\cos{(\theta_1)}}&{-u\sin{(\theta_1)}}\\
\end{array}\right)\left(\begin{array}{c}
{{\Delta\alpha}}\\
{{\Delta\tau}}\\
\end{array} \right)\; ,
\label{y'}
\ee
and
\begin{eqnarray}
D_2 &=&-b\left[2s\alpha_1\Delta\alpha +
u(2\tau_1-\tau_0)\Delta\tau\right]\nonumber\\
\eta &=& [u(\tau_1 - \tau_0) \cos{\theta_1} 
- b \alpha_1 \sin{\theta_1}] \Delta \tau - 
[s \alpha_1 \cos{\theta_1} + b \tau_1 \sin{\theta_1}] \Delta \alpha~~~\ .
\label{D2e}
\end{eqnarray}
Eqs. (\ref{D2e}) can be inverted to obtain $\Delta\alpha$
and $\Delta\tau$ as functions of $D_2$ and $\eta$. When the result is
inserted into Eq. (\ref{y'}), we obtain 
\be
\label{density}
\sigma(D_2,\eta)={b \over 2\sqrt{(b\tau_1)^2+(s\alpha_1)^2}}\,
{1\over u\tau_1|3\tau_0-4\tau_1|}\left(\tau_1({D_2\over b})^2 -
{us\tau_0\tau_1^2 \eta^2 \over (b\tau_1)^2+(s\alpha_1)^2} \right)~~~\ .
\ee
This implies
\be
D_2(\eta, \sigma)=
\sqrt{\frac{b^2}{1+(\frac{b^2}{us}-1)\frac{\tau_1}{\tau_0}}\eta^2
+
2b^2u|3\tau_0-4\tau_1|\sqrt{\frac{us}{b^2}\,(\tau_0-\tau_1)\tau_1+\tau_1^2}
\,\,\,\sigma}
~~~\ .
\ee
Along the $\hat{\sigma}$ direction ($\eta=0$), we have
\be
D_2(\sigma)= 2 C(\tau_1)\, b\sqrt{\sigma p}~~~\ ,
\label{D2sC}
\ee
with
\be
C(\tau_1)=\sqrt{ |3 -4 {\tau_1 \over \tau_0}| 
\sqrt{{{us}\over{b^2}}(1- {\tau_1 \over \tau_0}){\tau_1 \over\tau_0}
+({\tau_1 \over \tau_0})^2}}~~~~\ .
\ee
Combining Eqs. (\ref{km}) and (\ref{D2sC}), and minding the 
factor of two because two flows contribute, we 
have 
\be
d(\tau_1, \sigma) = {A(\tau_1) \over \sqrt{\sigma}} \Theta(\sigma)
\ee
with
\be
A(\tau_1)= \frac{d^2M}{d\Omega dt}
\frac{1}{bC(\tau_1)\sqrt{p}}\frac{\cos{(\alpha_1)}}{\rho(\tau_1)}~~~\ .    
\ee
In terms of $A_{0,n} = A_n(\tau_1 = \tau_0)$, for which estimates are
provided in Eq. (\ref{Anring}), we have
\be
A_n (\tau_1) = A_{0,n} {a_n \over \rho_n(\tau_1)} 
{\cos{\alpha_1(\tau_1)} \over C_n(\tau_1)}~~~~~\ .
\label{At1}
\ee
Note that $A(\tau_1)$ diverges at each of the three cusps because 
$C$ vanishes there.  In ref. \cite{sing} the caustic ring 
parameters ($a,~\tau,~b,~u,~s$) are related to the velocity 
distribution of the flow at last turnaround.

\subsubsection{Curvature radii everywhere}

The lensing by a caustic surface depends upon the curvature radius 
$R$ of the surface along the line of sight. $R$ is given by:
\be 
{1 \over R} = {(\cos{\omega})^2 \over R_1} + {(\sin{\omega})^2 \over R_2}
\label{Rad}
\ee
where $R_1$ and $R_2$ are the principal curvature radii of the
surface and $\omega$ is the angle between the line of sight and the
direction 
associated with $R_1$.  In this subsubsection, we adopt the convention 
that $R$ is positive (negative) if, along  the line of sight, the 
surface curves towards (away from) the side with two extra flows.  If
$R$ is positive, the surface is called `concave'; see Fig. 4.  If $R$ 
is negative, the surface is called `convex'; see Fig. 6.  It is a
straightforward exercise to calculate the curvature radii of the caustic
ring surface at an arbitrary point $(\alpha_1 (\tau_1), \tau_1)$.  We find
\be
R_1 (\tau_1) = -2 {\sqrt{su} \over b} p \sqrt{{\tau_1 \over \tau_0}}\,
|3 - 4 {\tau_1 \over \tau_0}|\, 
\left(1 - (1 - {b^2 \over su}){\tau_1 \over \tau_0}\right)^{3 \over 2}
\label{R1}
\ee
in the direction of the cross-sectional plane of the caustic ring, and
\be 
R_2 (\tau_1) = \pm {\rho_1 (\tau_1) \over \sin{\theta(\tau_1)}} =
\pm {\sqrt{su} \over b} \sqrt{{\tau_0 \over \tau_1} - 1 + {b^2 \over su}}
\left(a + {u \over 2}(\tau_0 - \tau_1)(\tau_0 - 2 \tau_1)\right)~~,
\label{R2}
\ee
in the direction perpendicular to the cross-sectional plane. 
In Eq. (\ref{R2}), the + sign pertains if 
$~0 \leq {\tau_1 \over \tau_0} \leq {3 \over 4}$, whereas the - sign 
pertains if $~{3 \over 4} \leq {\tau_1 \over \tau_0} \leq 1$.  $R_1$ 
is always negative except at the three cusps where it vanishes. $R_2$
diverges at the cusp in the $z=0$ plane.

For  $0 \leq {\tau_1 \over \tau_0} \leq {3 \over 4}$, there is 
a pair of lines of sight for which the curvature vanishes.  They 
are at angles:
\be
\omega = \pm \arctan{\sqrt{- {R_2 \over R_1}}}
\ee
relative to the cross-sectional plane.  Gravitational lensing by 
a fold of zero curvature is discussed in section IV.C.
 
\subsubsection{Density near a cusp}

In this subsubsection, we derive the dark matter density profile near 
a cusp. For the sake of
convenience, we choose the cusp in the $z=0$ plane 
at $\rho = a + p \equiv \rho_0$, where $\alpha = \tau = 0$.  Very 
close to the cusp, we may neglect the term of order $\tau^2$ in 
Eqs. (\ref{flow}). Hence
\be
z=b\,\alpha\tau\,\, ,\hskip 0.5cm
\rho=\rho_0 -u\tau_0 \tau -\frac{s}{2}\alpha^2\; .
\label{kem2}
\ee
The term of order $\alpha^2$ cannot be neglected. We define new
dimensionless quantities \cite{sing}:
\be
A\equiv\frac{\alpha}{\tau_0}\sqrt{\frac{s}{u}}\,\, ,\hskip 0.5 cm
T\equiv\frac{\tau}{\tau_0}\,\, ,\hskip 0.5 cm
X\equiv{{\rho-\rho_0}\over p}\,\,
,\hskip 0.5 cm
Z\equiv\frac{z\sqrt{\zeta}}{p}\; ,
\label{capvar}
\ee
where $\zeta = {su \over b^2}$.  In terms of these, Eqs. (\ref{kem2}) 
become
\be
Z&=&2AT\;,\label{kemn1}\\
X&=&-2T-A^2\;.\label{kemn2}
\ee
The Jacobian, Eq. (\ref{jaco}), becomes:
\be
D_2(A,T)= 2bp\left(T-A^2\right)\; .
\label{dat}
\ee
Substitution of Eq. (\ref{kemn1}) into Eq. (\ref{kemn2}) yields the third 
order polynomial equation:
\be
A^3+X A+ Z=0\; .
\label{pol}
\ee
The discriminant is:
\be
\delta=\left(\frac{Z}{2}\right)^2+\left(\frac{X}{3}\right)^3\; .
\label{discr}
\ee
If $\delta>0$, the cubic equation has one real root, and two complex
roots which are complex conjugate of each other.  If $\delta < 0$, all 
the roots are real and unequal.  For $\delta=0$, all the roots are real
and at least two are equal. The number of real roots is the number of
flows at a given location. The tricusp has two flows outside and four 
inside.  However, in the neighborhood of a cusp, one of the flows of 
the tricusp is non-singular.  It does not participate in the cusp 
caustic.  To include the root corresponding to the non-singular flow 
near $(z, \rho) = (0, \rho_0)$, one must keep the term of order $\tau^2$ 
in Eq. (\ref{kem2}).

$\delta=0$ is the equation for the caustic surface in physical
space. Indeed, Eq. (\ref{dat}) implies that at the caustic
$T=A^2$. Therefore, $Z=2T^{3/2}$ and $X=-3T$. A straightforward
calculation shows that 
\be
\delta=\frac{D_2}{54\,bp}(A^4+7A^2T-8T^2)\; .
\ee
Thus $D_2=0$ implies $\delta=0$. However, the converse is not true:
$\delta=0$ does not imply $D_2=0$ because not all flows at the 
location of the caustic surface are singular. 

Eq. (\ref{km}) for the density becomes
\be
d = {1 \over 2 \rho_0 b p} {{d^2M}\over{d\Omega dt}}
\sum_{j=1}^n {1 \over |T - A^2|_j}\; ,
\label{dTA}
\ee
where the sum is over the flows, i.e. the real roots of the cubic
polynomial (\ref{pol}).  If $\delta>0$, the one real root is:
\be
A=(-{Z\over 2}+\sqrt{\delta})^{1/3}+(-{Z\over 2}-\sqrt{\delta})^{1/3}.  
\label{A1}
\ee
It describes the one flow outside the cusp.  Using
Eqs. (\ref{kemn2},~~\ref{A1}) in Eq. (\ref{dTA}) we obtain
\be
d=\frac{1}{\rho_0\, b\, p}\frac{d^2M}{d\Omega dt}
\frac{1}{|X-3(-\frac{Z}{2}+\sqrt{\delta})^{2/3}
-3(\frac{Z}{2}+\sqrt{\delta})^{2/3}|} ~~~\ .
\ee
Just above or below the cusp, where $X=0$ and $|Z|\ll 1$, we have 
\be
d=\frac{1}{3\, b\, p\, \rho_0}\frac{d^2M}{d\Omega
dt}\frac{1}{|Z|^{2/3}}~~~\ .
\ee
On the other hand, if we approach the cusp in the plane of the ring from
the outside, where $Z=0$ and $0<X\ll 1$, we find  
\be
d=\frac{1}{b\, \rho_0}\frac{d^2M}{d\Omega
dt}\frac{1}{(\rho-\rho_0)}~~~\ . 
\label{fir}
\ee
Next, we calculate the density inside the cusp, where $\delta <0$.
The three real roots of the polynomial (\ref{pol}) are:
\be
A_1&=&2\sqrt{\frac{-X}{3}}\cos{\theta}\\
A_2&=&2\sqrt{\frac{-X}{3}}\cos{(\theta +\frac{2\pi}{3})}\\
A_3&=&2\sqrt{\frac{-X}{3}}\cos{(\theta +\frac{4\pi}{3})}\; ,
\ee
where $\cos{3\theta}\equiv -\frac{Z}{2}\left(-\frac{3}{X}\right)^{3/2}$
and $0\leq\theta\leq{\pi\over 3}$.  Inserting them into Eq. (\ref{dTA}) 
and using Eq. (\ref{kemn2}), we obtain
\be
d=\frac{1}{b\, p\, \rho_0}\frac{d^2M}{d\Omega
dt}\left(\frac{1}{-X}\right)\left(\frac{1}{4\cos^2(\theta)-1}
+\frac{1}{4\cos^2(\theta+\frac{2\pi}{3})-1}+
\frac{1}{1-4\cos^2(\theta+\frac{4\pi}{3})}\right)~~~\ ,
\ee
where each term in the parentheses corresponds to one of the three
flows. Adding the individual flow densities yields: 
\be
d=\frac{2}{b\, p\, \rho_0}\frac{d^2M}{d\Omega
dt}\frac{1}{|X|}\frac{1}{(\sqrt{3}-\tan{\theta})
\sin{2\theta}}~~~\ .
\ee
If we approach the cusp in the galactic plane from the inside, where
$Z=0$ and $0< -X \ll 1$, we have
\be
d=\frac{2}{b\, \rho_0}\frac{d^2M}{d\Omega
dt}\frac{1}{(\rho_0 -\rho)}~~~~,
\ee
which is twice the result in Eq. (\ref{fir}).  The gravitational 
lensing properties of a cusp are derived, for a special line of 
sight, in subsection IV.D.

\section{General formalism}
\label{sec:gf}

The first part of this section gives a brief account of the gravitational
lensing formalism \cite{lens}. In the second part we show how this
formalism can be streamlined in the special case of lensing by dark 
matter caustics.

In linear approximation, the deflection angle $\vec\theta$ of a light 
ray due to a gravitational field is given by:  
\be
\vec\theta =\vec{\nabla} {2\over{c^2}}\int \Phi~dy
\label{v}
\ee
where $\Phi$ is the Newtonian potential. We choose the $y$-axis 
in the direction of propagation of light. The deflection angle
$\vec{\theta}$ is related to the angular shift $~\vec{\xi_I}-\vec{\xi_S}~$
on the sky of the apparent direction of a source:
\be
\vec\theta
(\vec{\xi_{\tiny I}})={{D_S}\over{D_{LS}}}(\vec{\xi_I}-\vec{\xi_S})
\label{fund}
\ee
where $D_S$ and $D_{LS}$ are the distances of the source to the 
observer and to the lens respectively. ${\vec{\xi}}_{\tiny S}$ is the
angular position of the source in the absence of the lens whereas
${\vec{\xi}}_{\tiny I}$ is the angular position of the image with 
the lens present. The angles carry components in the $x$ and
$z$-directions: $\vec{\theta}= (\theta_x , \theta_z)$,
$\vec{\xi} = (\xi_x,\xi_z)$, etc.  Unless otherwise stated, 
we mean by a vector a quantity with 
components in the $x$- and $z$-directions.    

It is convenient to introduce a 2-D potential 
\be
\psi({\vec{\xi}}_{\tiny I})=\frac{2D_{LS}}{c^2\, D_L\, D_S}\int
dy~\Phi~~~\ ,
\ee
so that
\be
\vec\theta={D_{\tiny S} \over D_{\tiny LS}}
\vec{\nabla}_{\xi_{\tiny I}}\psi({\vec{\xi}}_{\tiny I})~~~\ ,
\label{th}
\ee
where $\vec{\nabla}_{{\xi}_{\tiny I}} = D_L\vec{\nabla}$. $D_L$ is
the distance of the observer to the lens. Eq. (\ref{fund}) then becomes:
\be
\vec{\xi}_{\tiny I}=\vec{\xi}_{\tiny
S}+\vec{\nabla}_{\xi_{\tiny I}}\psi(\vec{\xi}_{\tiny I})~~~\ .
\label{lens}
\ee
It gives the map $\vec{\xi}_{\tiny S}(\vec{\xi_{\tiny I}})$ from the image
plane to the source plane.  The inverse map may be one to one, or one to
many. In the latter case, there are multiple images and infinite
magnification when a pair of images merges. 

Our problem is to find the image map of a point source when the matter
distribution is given. The 2-D gravitational potential $\psi$ obeys the
Poisson equation:
\be
\nabla^2_{\xi_{\tiny I}}\psi={{8\pi G}\over{c^2}}{{D_L
D_{LS}}\over{D_{S}}}\Sigma= 2 {{\Sigma}\over{\Sigma_c}}~~~~\ ,
\label{Poisson}
\ee
where $\Sigma(\xi_{\tiny Ix}, \xi_{\tiny Iz})$ is the column density, 
i.e. the integral of the volume density along the line of sight:
\be
\Sigma(\vec{\xi}_{\tiny I})=\int dy\, d(D_L\xi_{\tiny
Ix},y,D_L\xi_{\tiny Iz})~~~\ ,
\label{S}
\ee
and $\Sigma_c$ is the critical surface density:
\be
\Sigma_c={{c^2D_{S}} \over{4\pi G D_L D_{LS}}}=0.347 \,
{\rm{g/cm^2}} \left({{D_{S}}\over{D_L D_{LS}}}\,{\rm{Gpc}}  
\right)\, .
\label{scri}
\ee
It is such that a uniform sheet of density $\Sigma_c$ focuses 
radiation from the source to the observer. 

Eq. (\ref{Poisson}) is solved by:
\be
\psi(\vec{\xi_{\tiny I}})
=\frac{1}{\pi\Sigma_c}\int~d^2\xi'_{\tiny I}~
\Sigma(\vec{\xi}'_{\tiny I})
\ln{|\vec{\xi}_{\tiny I}-\vec{\xi}'_{\tiny I}|}~~~~~~\ ,
\label{psi}
\ee
and hence 
\be
\Delta\vec{\xi}\equiv \vec{\xi}_{\tiny I}-\vec{\xi}_{\tiny S}
=\vec{\nabla}_{\xi_I} \psi (\vec{\xi}_{\tiny I})
={1\over{\pi\Sigma_c}}\int~d^2\xi'_{\tiny I}\, 
\Sigma(\vec{\xi}'_{\tiny I}){{\vec{\xi}_{\tiny I}-\vec{\xi}'_{\tiny I}}
\over{(\vec{\xi}_{\tiny I}-\vec{\xi}'_{\tiny I})^2}} ~~~\ .
\label{shift}
\ee 
The image structure, distortion and magnification are given by the
Jacobian matrix of the map $\vec\xi_S(\vec\xi_I)$ from image to 
source:
\be
K_{ij}\equiv {{\partial \xi_{\tiny Si}}\over{\partial
\xi_{\tiny Ij}}}=\delta_{ij}-\psi_{ij}
\label{K}
\ee
where $\psi_{ij}\equiv\frac{\partial^2\psi}
{\partial \xi_{\tiny Ii}\partial\xi_{\tiny Ij}}$. Because 
gravitational lensing does not change surface brightness, the
magnification ${\mathcal{M}}$ is the ratio of image area to source
area. Therefore
\be
{\mathcal{M}}={1\over{|\det{(K_{ij})}|}}~~~\ .
\ee
To first order, for $\psi_{ij} \ll 1$,
\be
{\mathcal{M}}=1+\nabla_{\xi_I}^2\psi = 1+2{\Sigma\over\Sigma_c} ~~~\ .
\label{mag}
\ee
To get the largest lensing effects, we wish to minimize $\Sigma_c$, 
given in Eq. (\ref{scri}).  For fixed $D_S$, the minimum occurs when 
the lens is situated half way between the source and the observer. 
Also $D_S$ should be as large as possible.  For our estimates, we 
will assume that the source is at cosmological distances, e.g. 
$2D_L=2D_{LS}=D_S = 1 {\rm Gpc}$, in which case 
$\Sigma_c = 1.39$ g/cm$^2$. 

For a general mass distribution, the gravitational lensing effects
are obtained by first calculating the column density, Eq. (\ref{S}), 
and then the image shift, Eq. (\ref{shift}).  This procedure can be  
simplified when the lens is a dark matter caustic.  We are interested 
in lines of sight which are tangent to a caustic surface, because the 
column density $\Sigma$ has the highest contrast there.  We assume that,
in the neighborhood of the tangent point, the flow is independent of 
$y$ except for a shift $\vec{x}(y)$ of the caustic surface with $y$. 
The density can then be written as:
\be
d(x,y,z)=d(x-x(y),z-z(y))~~~\ ,
\label{den}
\ee
where $d(x,z)$ is the density of the 2-D flow in the plane orthogonal 
to the line of sight. The flow in the $y$-direction is irrelevant
because lensing depends only on the column density $\Sigma$. 

In the limit of zero velocity dispersion, a 2-D flow is specified by 
giving the spatial coordinates $\vec{x}(\alpha,\beta,t)$ of the particle
labeled $(\alpha,\beta)$ at time $t$, for all $(\alpha,\beta,t)$. The
labels $\alpha$ and $\beta$ are chosen arbitrarily. At a given time $t$,
$\vec{x}(\alpha,\beta,t)=\vec{x}$ has a discrete number of solutions
$(\alpha, \beta)_j(\vec{x},t)$ labeled by $j=1, ..., n(\vec{x}, t)$. 
$n$ is the number of distinct flows at $(\vec{x}, t)$. Here, $t$ is the
time at which the light ray passes by the caustic. Henceforth we will 
not show the time dependence explicitly. The particle density in
physical space is: 
\be
d(x, z)=\sum_{j=1}^{n(\vec{x})}~{{d^2 \Lambda}\over{d\alpha d\beta}}~
{1\over{{|{{\dd (x, z)}\over{\dd (\alpha,\beta)}}|}}}
\Bigg|_{(\alpha,\beta)={(\alpha,\beta)}_j(\vec{x})} ~~~\  .
\label{dxi}
\ee
where $\Lambda$ is the mass per unit length in the direction of the 
line of sight and  ${{d^2 \Lambda}\over{d\alpha d\beta}}$ is the 
$\Lambda$ density in parameter space.  Inserting Eqs. (\ref{S}) 
and (\ref{den}) into Eq. (\ref{shift}) we obtain: 
\be
\vec{\nabla}_{\xi_I}\psi(\vec{\xi_{\tiny I}}=\vec{x}/D_L)=
\frac{1}{\pi\Sigma_c}\int dy\int d^2\xi'_{\tiny I}~
{{\vec\xi_{\tiny I}-\vec\xi'_{\tiny I}}\over
{|\vec\xi_{\tiny I} -\vec\xi'_{\tiny I}|^2}}~
d(D_L\xi'_{\tiny Ix} -x(y), D_L\xi'_{\tiny Iz} -z(y)) \hskip 0.2 cm .
\label{np}
\ee
When $d$ is replaced by the caustic density (\ref{dxi}),
Eq.(\ref{np}) becomes:
\begin{eqnarray}
\vec{\nabla}_{\xi_I}\psi (\vec{\xi_{\tiny I}}=\vec{x}/D_L) &=& \nonumber\\
\frac{1}{\pi\Sigma_c D_L}\int dy \int dx'\,dz'&&
{{\vec{x}-\vec{x}'}\over{(\vec{x}-\vec{x}')^2}}~
\sum_{j=1}^n {1\over{{|{{\dd (x', z')}\over{\dd(\alpha,\beta)}}|}}}
\frac{d^2 \Lambda}{d\alpha d\beta}
\Bigg|_{(\alpha,\beta)={(\alpha,\beta)}_j(\vec{x}' -\vec{x}(y))}~~~\ .
\label{ara}
\end{eqnarray}
Changing variables from $(x',z')$ to $(\alpha,\beta)$ and assuming 
that the density in parameter space varies only slowly over the region 
of integration, we rewrite Eq. (\ref{ara}) as:
\be
\vec{\nabla}_{\xi_I}\psi (\vec{\xi_{\tiny I}}=\vec{x}/D_L)=
\frac{1}{\pi\Sigma_c D_L}
\frac{d^2 \Lambda}{d\alpha d\beta}\int dy\int d\alpha\, d\beta\; 
\frac{\vec{x}-\vec{x}(\alpha,\beta)-\vec{x}(y)}
{|\vec{x}-\vec{x}(\alpha,\beta)-\vec{x}(y)|^2} ~~~\ .
\label{gradpot}
\ee
Further simplification is achieved by defining the complex integral:
\be
I(x, z)\equiv\int d\alpha d\beta~\frac{1}{x-x (\alpha ,\beta)
+i(z-z(\alpha , \beta))}~~~~~\ ,
\label{I}
\ee
in terms of which the shifts are given by:
\be
\Delta \xi_{\tiny x}&=& 
\frac{\partial\psi}{\partial\xi_{\tiny Ix}}=
\frac{1}{\pi\Sigma_c D_L}\frac{d^2 \Lambda}{d\alpha d\beta}
~\mbox{Re}\,\int dy\,\, I(\vec{x}-\vec{x}(y))~~~\ ,\nonumber\\
\Delta \xi_{\tiny z}&=& 
\frac{\partial\psi}{\partial\xi_{\tiny Iz}} =
-\frac{1}{\pi\Sigma_c D_L}\frac{d^2 \Lambda}{d\alpha d\beta}
~\mbox{Im}\,\int dy\,\, I(\vec{x}-\vec{x}(y))~~~\ .
\label{LCI1}
\ee
Eqs. (\ref{I}) and (\ref{LCI1}) are useful when the caustic has 
contrast in the two dimensions transverse to the line of sight, 
e.g. near a cusp.

However, in many applications, the caustic has contrast in only one
of the dimensions transverse to the line of sight.  Choosing $\hat x$ 
to be the trivial direction, Eq. (\ref{Poisson}) is reduced to:
\be
{{d^2\psi}\over{d\xi^2_{\tiny I}}}(\xi_{\tiny I})=
{2 \over \Sigma_c}~\Sigma(z=D_L\xi_{\tiny I})~~~\ ,
\label{Poi2}
\ee  
and the column density is given by:
\be
\Sigma(\xi_{\tiny I})=\int~dy~d(z-z(y))~~~\ .
\label{col1}
\ee
The flow is now effectively one dimensional.  Its physical space 
density is given by 
\be
d(z)=\sum_{j=1}^{n(z)}~{{d \Lambda}\over{d\alpha}}~
{1\over{{|{{dz}\over{d\alpha}}|}}}\Bigg|_{\alpha={\alpha}_j(z)}~~~\ ,
\label{d1D}
\ee
where $\Lambda$ is the mass surface density in the two trivial 
directions ($x$ and $y$) and ${d\Lambda \over d\alpha}$ is the 
$\Lambda$ density in parameter space.  The 1-D Green's function 
is $G={1 \over 2}(|\xi|+a\xi)+b$, where $a$ and $b$ are arbitrary
constants. 
The shift is:
\be
{{d\psi}\over{d\xi_{\tiny I}}}(\xi_{\tiny I})= {1\over{\Sigma_c}}\int
d\xi'_{\tiny I}~\Sigma(\xi'_{\tiny I})~
({\rm Sign}(\xi_{\tiny I}-\xi'_{\tiny I})+a)~~~\ .
\label{sign}
\ee
The constant $a$ causes an overall shift of the image, which does not
concern us. We choose $a=-1$. Repeating the steps done earlier for the 
2-D case, Eq. (\ref{sign}) is re-expressed as:
\be
\Delta\xi =
-{{2}\over{\Sigma_c D_L}}{{d \Lambda}\over{d\alpha}}
\int dy \int d\alpha~\Theta (-z+z(\alpha)+z(y))~~~\ . 
\label{theeq}
\ee
In the next section Eq. (\ref{theeq}) is used to calculate the shifts 
due to simple folds of dark matter flows, and Eq. (\ref{LCI1}) for the
shifts due to a cusp.

\section{Applications}
\label{sec:fc}

\subsection{Lensing by a concave fold}
\label{subsection:sccf}

We consider a simple fold caustic which has curvature radius $R$ along 
the line of sight.  We assume that the surface is concave, i.e. it is
curved in the direction of the two extra flows.  See Fig. \ref{fig:concave}.
The outer caustics of dark matter halos are examples of concave caustic 
surfaces.  The convex case is discussed in the next subsection.

In the neighborhood of the point $P$ of closest approach of the line of
sight with the caustic surface, we choose coordinates such that $\hat{z}$
is perpendicular to the surface whereas $\hat{x}$ and the direction
$\hat{y}$ of the line of sight are parallel. $P$ has coordinates
$x=y=z=0$. In the $y=0$ plane, the flow is given by 
$z(\alpha)=-{1\over 2}h\alpha^2$, where $h$ is a positive constant.
Using Eq. (\ref{d1D}) we find the density in the $y=0$ plane:
\be
d(\sigma) = A~{\Theta(\sigma) \over \sqrt{\sigma}}~~~\ ,
\label{Ah}
\ee
where $\sigma = -z$, and 
\be
A=\sqrt{{2\over h}}~{{d \Lambda}\over{d\alpha}}~~~\ .
\label{Aa}
\ee
For $y \neq 0$, the density is still given by Eq. (\ref{Ah})
but with $\sigma=z(y)-z$ and $z(y)=-\frac{y^2}{2R}$.
We calculate the shift using Eq. (\ref{theeq}):
\begin{eqnarray}
\Delta\xi&=&-\frac{2}{\Sigma_c D_L}
\frac{d \Lambda}{d\alpha}\int_{-\sqrt{-2Rz}}^{\sqrt{-2Rz}} dy\int d\alpha~
\Theta \left(-z - {h \over 2} \alpha^2 - {y^2 \over 2R} \right)\nonumber\\
&=& {{4\pi}\over{\Sigma_c}}{{d\Lambda}\over{d\alpha}}\sqrt{{R\over h}}\,
\Theta (-\xi_{\tiny I})\,\xi_{\tiny I}~~~\ .
\label{grt2}
\end{eqnarray} 
Hence
\be
\Delta\xi =\xi_{\tiny I}-\xi_{\tiny S} =
\eta\,\xi_{\tiny I}\,\Theta{(-\xi_{\tiny I})}~~~\ ,
\label{Dks}
\ee
with
\be
\eta = {{2\pi A\sqrt{2R}}\over{\Sigma_c}}~~~\ .
\label{ep}
\ee
One can also obtain this result by calculating the column density: 
\be
\Sigma(\xi_{\tiny I})=\int dy~d(y,z=\xi_{\tiny I})
=\int dy~\frac{A~\Theta(-\xi_{\tiny I}D_L-{{y^2}\over{2R}})}
{\sqrt{-\xi_{\tiny I}D_L -{{y^2}\over{2R}}}}=
\pi A\sqrt{2R}\,\Theta (-\xi_{\tiny I}) ~~~\ ,
\label{cofcod}
\ee
and solving Eq. (\ref{Poisson}): 
\be
\xi_{\tiny I}-\xi_{\tiny S}={{d\psi}\over{d\xi_{\tiny I}}}
={2\over{\Sigma_c}}\int d\xi_{\tiny I}\,\Sigma(\xi_{\tiny I})
=\eta\,\xi_{\tiny I}\,\Theta(-\xi_{\tiny I}) ~~~\ .  
\label{-}
\ee
Eqs. (\ref{cofcod}, \ref{-}) were first obtained by Hogan \cite{Hogan}.
The agreement with Eq. (\ref{Dks}) validates the formalism derived in
section \ref{sec:gf}.  Fig. \ref{fig:shiftconcave} plots the source
position $\xi_S$ versus the image position $\xi_{\tiny I}$.

The magnification is:
\be
{\mathcal{M}}={{d\xi_{\tiny I}}\over{d\xi_{S}}}
= 1 + \eta\;\Theta{(-\xi_I)} + 0(\eta^2)~~~\ .
\label{Mm}
\ee
When the line of sight of a moving source crosses the surface of a
simple concave fold, the component of its apparent velocity perpendicular
to the fold changes abruptly.  Also, a discontinuity occurs in the
magnification of the image.  Both effects are of order $\eta$.

We estimate $\eta$ for the outer caustics of galactic halos
using Eqs. (\ref{AN} - \ref{Fne}):
\be
\{\eta_n=\frac{\sqrt{2}\, v_{\tiny{\rm rot}}^2}{G\,\Sigma_c}\,
\frac{F_n}{R_n} \, : n=1, 2, ...\}\sim
(7,~6,~6,~6,~6,\,...)\cdot 10^{-3}\cdot\nonumber\\
\cdot \left(\frac{D_L D_{LS}}{D_S~{\rm Gpc}}\right)
\left(\frac{v_{\rm \tiny{rot}}}{220 {\rm km/s}}\right)
\left(\frac{h}{0.7}\right)~~~\ .
\label{epe}
\ee
A magification of order $10^{-2}$ seems hard to observe. However, 
the images of extended sources may be modified in recognizable
ways. In particular, straight jets would be seen with an abrupt 
bend where their line of sight crosses a fold. Indeed the image
is stretched by the factor $1+\eta$ in the direction perpendicular
to the caustic, on the side with the two extra flows.  If the jet makes
angle $\alpha$ with the normal, it appears bent by an angle
$\delta \equiv \frac{1}{2}\eta\sin{\left({2\alpha}\right)}$. 
Searching the sky for bends in extended sources may be a realistic
approach to detecting caustic structure in galactic halos \cite{Hogan}.  

\subsection{Lensing by a convex fold}
\label{subsection:scvf}

By definition, a convex fold is curved in the direction opposite to the
side with two extra flows. See Fig. \ref{fig:convex}. Using the
conventions of the previous subsection, we write the equation 
for the displacement of the surface along the line of sight
as $z(y)={{y^2}\over{2R}}~$, and that for the flow as
$~z(\alpha)=-{1\over 2}h\alpha^2~$, for small $z$, with $R, h>0$. 
Eqs. (\ref{Ah}) and (\ref{Aa}) still apply, with 
$~\sigma = {y^2/2R} - z~$.  Eq. (\ref{theeq}) yields the shift: 
\be
\Delta\xi =-{{8}\over{{\Sigma_c}D_L}}{{d \Lambda}\over{d\alpha}}
\left\{\Theta(-\xi_{\tiny I})\int_0^\infty dy 
+\Theta(\xi_{\tiny I})\int_{\sqrt{2RD_L\xi_{\tiny I}}}^\infty dy\right\}
\sqrt{{2 \over h}\left({{y^2}\over{2R}}-D_L\xi_{\tiny I}\right)} ~~~\ .
\ee
We introduce a cut-off $L$ for the integral over large $y$.  $L$ can 
be thought of as the length scale beyond which our description of the 
flow is invalid.  The above equation becomes:
\be
\Delta\xi =-\frac{4}{\Sigma_c D_L}
\frac{d \Lambda}{d\alpha}\frac{1}{\sqrt{hR}}
\left\{L\sqrt{L^2-2RD_L\xi_{\tiny I}}
-2RD_L\xi_{\tiny I}\,\ln{\left(\frac{L +\sqrt{L^2 -2RD_L\xi_{\tiny I}}}
{\sqrt{2RD_L|\xi_{\tiny I}|}}\right)}\right\}.
\label{Dxiz}
\ee
By expanding Eq. (\ref{Dxiz}) in powers of ${1\over L}$, and using
Eq. (\ref{Aa}), we obtain the $\xi_{\tiny I}$-dependent shift:
\be
\Delta\xi=\xi_{\tiny I} -\xi_{\tiny S}
=-\frac{\eta^\prime}{\pi}\left[\,\ln{\left({{RD_L|\xi_{\tiny I}|}
\over{2 L^2}}\right)}-1\right]\xi_{\tiny I} ~~~\ ,
\label{shifc}
\ee  
where $\eta^\prime$ is [like $\eta$ in Eq. (\ref{ep})] given by
\be 
\eta^\prime = {2 \pi A \sqrt{2 R} \over \Sigma_c}~~~\ .
\label{epp}
\ee
The magnification is :
\be
{\mathcal{M}}=\frac{d\xi_{\tiny I}}{d\xi_{\tiny S}}=
\left|1 + \frac{\eta^\prime}{\pi}~
\ln{\left({{R\,D_L\,|\xi_{\tiny I}|}
\over{2 L^2}}\right)}\right|^{-1} ~~~\ .
\label{magc}
\ee
The cut-off $L$ has an effect on both the magnification and the
elongation of the image in the direction normal to the caustic surface,
but that effect is $\xi_I$ independent.  $L$ has a global effect on the
image, as opposed to an effect localized near $\xi_I = 0$. 

Fig. \ref{fig:shiftconvex} plots $\xi_{\tiny S}$ versus $\xi_{\tiny I}$.
It shows that a convex fold can cause a triple image of a point
source.  In particular, when the source is exactly behind the 
caustic ($\xi_S = 0$), the images are at $\xi_I = -\xi_c$, 0, and 
$+\xi_c$, with 
\be
\xi_c = {2 L^2 \over R D_L} \exp{(-{\pi \over \eta^\prime} + 1)}~~~\ .
\label{xic}
\ee
Sufficiently far from the caustic, a point source has a single 
image, say at $\xi_{\tiny I1}>0$. When the line of sight approaches the
caustic surface tangent point,  two new images appear on top of each other
at 
$\xi_{\tiny I2}=\xi_{\tiny I3} = -\xi_c/e$. At that moment, the
magnification at $\xi_{I2}$ is infinite, and $\xi_S = 
\eta^\prime \xi_c/e \pi$.  As the source crosses the caustic,
$\xi_{\tiny I2}$ moves towards $\xi_{\tiny I1}$ and finally
merges with it. When $\xi_{\tiny I1} = \xi_{\tiny I2} = +\xi_c/e$, 
the magnification diverges again. After that, only the image at
$\xi_{\tiny I3}$ remains.     

Let us apply the above results to the line of sight in the plane 
of a caustic ring at the sample point, $(z=0, \rho=a)$, discussed 
in subsection II.B.1. Setting $R = a_n$ and using Eq. (\ref{Anv2}) 
we obtain for the $n$th ring
\be
\eta^\prime_n =\frac{2\pi\, A_{0,n}\sqrt{2a_n}}{\Sigma_c}=
\frac{v^2_{\tiny {\rm rot}}}{\sqrt{2} G\Sigma_c}
\frac{f_n}{\sqrt{a_n\,p_n}}\frac{v_n}{b_n} ~~~\ .
\ee
Using Eq. (\ref{Anring}) to estimate $A_{0,n}$, we find 
\be
\{\eta^\prime_n: n=1,2,...\}\sim (5,~4,~4,~4,~4,...)
\cdot 10^{-2}~\frac{D_L\,D_{LS}}{D_S\,{\rm Gpc}}\nonumber\\
\cdot\left(\frac{0.27}{j_{\rm\tiny{max}}}\right)
\left(\frac{h}{0.7}\right)
\left(\frac{v_{\tiny{\rm rot}}}{220\,{\rm km/s}}\right)~~~\ .
\ee
For such small values of $\eta^\prime$ the angular distance 
between the triple images is exponentially small and unresolvable 
with present and foreseeable instruments.

The image of an extended object is stretched in the direction
perpendicular to the caustic by the relative amount
\be 
{\mathcal{M}} - 1 = - {\eta^\prime \over \pi} 
\ln{\left({R D_L |\xi_I| \over 2 L^2}\right)}~~~\ .
\label{Str}
\ee
The image is stretched for $\xi_I < \xi_d$, and compressed for 
$\xi_I > \xi_d$, where
\be
\xi_d = {2 L^2 \over R D_L}~~~\ .
\label{xid}
\ee
In the case of the sample point $(z=0, \rho=a)$ with line of sight 
in the $z=0$ plane, the cut-off length $L$, i.e. the distance in the 
$y$-direction over which our desciption of the flow is valid, is of 
order $\sqrt{2 a p}$.  In that case $\xi_d \sim 4 p/D_L$.  Since 
$p/D_L$ is the transverse angular size of the caustic ring, our 
description certainly fails for $\xi_I > p/D_L$.  So, over the region 
where our calculation is valid, the image is magnified.  The effects are
generically of 
order one percent, increasing to several percent when 
$\xi_I \sim 10^{-3} \xi_d$, for caustic rings at cosmological
distances.

\subsection{Lensing by a fold with zero curvature}

We saw in section II.B.3 that the surface of a caustic ring has tangent
lines along which the curvature vanishes.  One may speculate that the
lensing effects of a caustic surface are strongest when the line of sight
is tangent to the surface in a direction of zero curvature, because the
line of sight stays close to the caustic over greater depths in that case.  
If the line of sight of an observer looking at the outside profile of a
caustic ring is, at some point on the profile, in a direction of zero
curvature, then the equation for the intersection of the caustic surface
with the plane containing the outward normal to the surface ($\hat{z}$)
and the line of sight ($\hat{y}$) is $z(y)=-\frac{y^4}{4U}$ where $U$ is
positive and has dimensions of (length)$^3$.  In such a case, the cubic
term in the Taylor expansion of $z(y)$ is absent because the line of sight
remains everywhere outside the caustic ring tube.  We did not calculate
$U$ for caustic rings but expect $U \sim$ (kpc)$^3$ in order of magnitude.  
The flow is given by $z(\alpha)=-{1\over 2}h\alpha^2$ with $h>0$, as
before, and Eq. (\ref{Aa}) holds.

Using the above expressions for $z(y)$ and $z(\alpha)$ in 
Eq. (\ref{theeq}), we find the shift
\be
\Delta\xi &=&-{2 \over \Sigma_c D_L}{d \Lambda \over d\alpha}
\int dy \int d\alpha~
\Theta\left(-z-{1 \over 2}h\alpha^2-{y^4 \over 4U}\right)\nonumber\\
&=&-{8 \over \Sigma_c D_L} {d \Lambda \over d\alpha}
{\left(-4Uz\right)^{3/4} \over \sqrt{2hU}}
\Theta(-\xi_I)\int_0^1 dt \sqrt{1-t^4}\nonumber\\
&=& -\Theta(-\xi_I) \left(-\xi_0\,\xi_I^3 \right)^{1/4}~~~~\ ,
\label{dxio}
\ee
where 
\be 
\xi_0 = (9.89~\frac{A}{\Sigma_c})^4 \frac{U}{D_L}~~~\ .
\label{xi0}
\ee
The magnification is:
\be
{\mathcal{M}} =  \left|1 - {3 \over 4}\Theta(-\xi_I)
\left(-\frac{\xi_0}{\xi_I}\right)^{1/4}\right|^{-1} ~~~\ .
\ee
Fig. \ref{fig:shiftflat} shows $\xi_{\tiny S}$ versus $\xi_{\tiny I}$.
Triple images occur when $|\xi_I| \lesssim \xi_0$. Unfortunately, for the
zero curvature tangents of caustic rings envisaged above, $\xi_0$ is very
small.  For $2D_L=2D_{LS}=D_S= {\rm Gpc}$, 
$A = 3 \cdot 10^{-4} {{\rm gr} \over {\rm cm}^2 {\rm kpc}^{1 \over 2}}~$, 
and $U = ({\rm kpc})^3$, one finds $\xi_0 = 4 \cdot 10^{-17}$.  Hence the 
triple images cannot be resolved.  Also, even at angular distances 
as small as $\xi_I \sim 10^{-9}$, the magnification and image 
distortion is only of order 1\%. 

\subsection{Lensing by a cusp}

In this subsection, we investigate a line of sight parallel to the
plane ($z=0$) of a caustic ring, and passing near the cusp at $\rho =
\rho_0$.  See Fig. \ref{fig:tricusp2}.  We use the 2-D lensing equations
derived in section \ref{sec:gf}. The shifts are given by Eq. (\ref{LCI1}) 
in terms of the complex integral $I$ of Eq. (\ref{I}). Using
Eq. (\ref{kem2}), we have 
\be
I=\int d\alpha\int d\tau \frac{1}{\rho-a -\frac{1}{2}u\tau_0^2
+u\tau_0\tau+\frac{1}{2}s\alpha^2+i(z-b\alpha\tau)} ~~~\ .
\label{Iint}
\ee
Because we are close to the cusp, the term of order $\tau^2$ is 
neglected in the denominator of the integrand.  In terms of the
parameters defined in Eqs. (\ref{capvar}), 
\be
I=\frac{2}{b\sqrt{\zeta}}\int dT\int_{-\infty}^{\infty}
\frac{dA}{A^2- 2 i {AT \over \sqrt{\zeta}} +2T + X + 
i\frac{Z}{\sqrt{\zeta}}}~~~~\ .
\label{reint}
\ee
The integration over $A$ yields
\be
I=\frac{2\pi}{b\sqrt{\zeta}}\int_{\frac{|T|}{\sqrt{\zeta}}<C}\,
\frac{dT}{\sqrt{\frac{T^2}{\zeta}+2T+X+i\frac{Z}{\sqrt{\zeta}}}}~~~~,
\label{IASZ}
\ee
where\hskip 0.3 cm 
$C\equiv{\rm Re}\,\,\sqrt{\frac{T^2}{\zeta}+2T+X+i\frac{Z}{\sqrt{\zeta}}}
>0\,$. Near the cusp ($X, Z << 1$) the terms of order $T^2$ can be
neglected since they are unimportant in the denominator of
Eq. (\ref{reint}).  Eq. (\ref{IASZ}) becomes then:
\be
I(X,Z)=\frac{2\pi}{b\sqrt{\zeta}}\int\frac{dT}{\sqrt{2T+X
+i\frac{Z}{\sqrt{\zeta}}}}~~~ ,
\label{2pi}
\ee
where the integration domain is defined by the inequality:
\be
T^4-\zeta (2T^3+XT^2)-{{\zeta Z^2}\over{4}}<0~~~~\ .
\label{relin}
\ee
Let us call $T_i\,$ $(i = 1,..., 4)$ the roots of the polynomial on the
left hand side of (\ref{relin}). Near the cusp, three of the roots $T_i$
are near zero and one of the roots is close to $2\zeta$. Let us call the 
latter $T_4$.  In order to find the roots near $T=0$ we neglect the
quartic term and solve the cubic
equation:
\be
T^3+\frac{X}{2}T^2+\frac{Z^2}{8}=0~~~~\ .
\label{cubic}
\ee
This equation is also obtained by eliminating $A$ from
Eqs. (\ref{kemn1}) and (\ref{kemn2}). Hence the solutions of
Eq. (\ref{cubic}) are the
values of $T$ for the flows near the cusp. As was discussed in
section II.B.4 , the number of flows is determined by $\delta$, 
given in Eq. (\ref{discr}). Outside the cusp, where $\delta > 0$, 
there is a single root $T_1$. Inside the cusp  where $\delta < 0$, 
there are three roots $T_1$, $T_2$ and $T_3$. Fig. \ref{fig:quartic} 
shows the quartic polynomial, Eq. (\ref{relin}), for various values 
of $X$ and $Z$, and $\zeta = 1$.

For a line of sight just outside the cusp, the quartic polynomial is
negative from $T_1$ to $T_4$ with:
\be
T_1=-{1\over 2}\left(\sqrt{\delta}
-\frac{|Z|}{2}\,\,\right)^{2/3}
-{1\over 2}\left(\sqrt{\delta}              
+\frac{|Z|}{2}\,\,\right)^{2/3}
-{X\over 6}~~~~\ .
\label{T1}
\ee
Eq. (\ref{2pi}) becomes in that case:
\be
I=\frac{2\pi}{b\sqrt{\zeta}}\Bigg\{\sqrt{2T_4+X+i\frac{Z}{\sqrt{\zeta}}}
-\sqrt{2T_1+X+i\frac{Z}{\sqrt{\zeta}}}\,\Bigg\}~~~~\ .
\label{Ikoc}
\ee
Let us consider the line of sight defined by Fig. \ref{fig:tricusp2}. 
Eq. (\ref{LCI1}) gives the shift in the direction perpendicular to 
the plane of the cusp as
\be
\Delta \xi_{z} = - {1 \over \pi \Sigma_c D_L}
{d^2 \Lambda \over d\alpha d\tau}\; {\rm Im}
\int dy~I(X - X(y), Z) 
\label{ano}
\ee
where $X(y) = - {y^2 \over 2 \rho_0 p}$ is the shift of the cusp 
as a function of depth.  The integral in Eq. (\ref{ano}) with the
integrand given by Eqs. (\ref{T1}) and (\ref{Ikoc}) can be done 
numerically.  Here we only give a rough estimate, to determine 
the order of magnitude and qualitative properties of the lensing 
effects.  

For $X=0$ and $Z<<1$, we have 
\be
I=\frac{2\pi}{b}\left[2-\frac{i}{\sqrt{\zeta}} \mbox{Sign}(Z)
|Z|^{1/3}-\frac{1}{2\zeta}|Z|^{\frac{2}{3}}\right]+O(Z)~~~~\
.
\label{Ikocx0}
\ee
Because the scaling law $X^3\sim Z^2$ holds close to the cusp, 
we expect Eq. (\ref{Ikocx0}) to be valid as long as the shift 
in the $x$-direction $|\Delta X|\lesssim |Z|^{\frac{2}{3}}$.  Hence 
Eq. (\ref{Ikocx0}) provides a good estimate of $I(X - X(y), Z)$
over a depth $2 \Delta y$ with  
$\Delta y \sim \sqrt{2 \rho_0 p}\, |Z|^{1 \over 3}$.  Therefore
\be
\Delta \xi_z \sim {2 \over \Sigma_c D_L b} 
\sqrt{{2 \rho_0 p \over \zeta}} 
\frac{d^2\Lambda}{d\alpha d\tau}~|Z|^{2\over 3}~{\rm Sign}(Z)~~~~\ .
\label{dxiz2}
\ee
Since we are near $\alpha = 0$,
\be 
\frac{d^2\Lambda}{d\alpha d\tau}
= {1 \over 2 \pi \rho_0} {d^2 M \over d\alpha d\tau}
= {1 \over \rho_0} {d^2 M \over d\Omega dt}~~~~\ .
\ee
Using this and Eq. (\ref{A_nc}), we obtain
\be
\Delta\xi_{z} \sim \eta''\,|\xi_{Iz}|^{{2}\over {3}}\,{\rm
Sign}(\xi_{Iz})~~~~ ,
\ee
where
\be
\eta''\equiv\frac{2\sqrt{2\, a}}{\zeta^{{1}\over {6}}\Sigma_c
D_L^{{1}\over {3}}} \,p^{{1}\over {3}}\,A_{0}~~~~\ .
\label{ep''}
\ee
The contribution to the magnification from distortion in the
$z$-direction is therefore:
\be
{\mathcal{M}}_z \sim
\left|1-\eta''\,{{2}\over {3}}|\xi_{Iz}|^{-{{1}\over{3}}}\right|^{-1}~~~~\
.
\ee
Fig. \ref{fig:shiftcusp} plots $\xi_{Sz}$ versus $\xi_{Iz}$.  For 
$\xi_{Iz} \lesssim \xi_1 \sim \eta''^3$, a point source
has triple images.  For the $n$th caustic ring, $\eta''$ 
can be estimated using Eqs. (\ref{a_n}, \ref{Anring}), and 
assuming $p \simeq 0.1 a,~\zeta \sim 1$: 
\be
\{\eta''_n : n=1, 2, ...\}\sim (1.5,~ 1,~ 0.7,~ 0.6,~ 0.5 .....)\cdot
10^{-4}\frac{D_L^{{2}\over {3}}\,D_{LS}}{D_S\,{\rm
Gpc}^{{2}\over {3}}}~\cdot\nonumber\\
\cdot\left(\frac{0.27}{j_{\rm\tiny{max}}}\right)^{{2}\over {3}}
\left(\frac{h}{0.7}\right)^{{2}\over {3}}
\left(\frac{v_{\tiny{\rm rot}}}{220\,{\rm km/s}}\right)^{4\over 3}~~~\ .
\label{ep''est}
\ee
At cosmological distances, the typical angular separation between 
the images is of order $10^{-12}$.  Unfortunately, here again, the 
images are too close to be resolved. However, for an
angular distance $\xi_I\sim 10^{-9}$, ${\mathcal{M}}_z$
is of order 10\%, which may be observable.  The lensing properties 
of a cusp for other lines of sight can be calculated numerically 
using Eqs. (\ref{LCI1}) and (\ref{Iint}).  The latter equation 
is replaced by Eqs. (\ref{Ikoc}) and (\ref{T1}), and 
$T_4 = 2 \zeta$, if the line of sight is everywhere outside 
the region with two extra flows.

\section{Conclusions}
\label{sec:con}

The dark matter caustics present in the halos of isolated galaxies
were discussed in Section II.  They are of two types, outer and inner.
The outer caustics are topological spheres surrounding the galaxy and
labeled by $n = 1,2,3 ..$.  Each is a simple fold ($A_2$) catastrophe
located near where a dark matter outflow reaches its furthest distance 
from the galactic center before falling back in.  The inner caustics 
are rings in or near the galactic plane.  They are also labeled by 
$n = 1,2,3 ..$.  A caustic ring is a closed tube whose cross-section 
is an {\it elliptic umbilic} ($D_{-4}$) catastrophe.  The cross-section, 
shown in Fig. \ref{fig:tricusp}, is characterized by three cusps.  A
caustic ring is located near where the particles with the most angular
momentum in a given inflow are at their distance of closest approach to
the galactic center before falling back out.  In section II, we gave 
estimates of the radii and the fold coefficients of the outer caustics, 
based upon the self-similar infall model.  For the inner caustic rings, 
the curvature radii and fold coefficients depend on position on the 
surface of the ring caustic.  We derived formulae for these quantities 
as a function of position (parametrized by $\tau_1$) and of the five
caustic ring parameters ($a,~b,~u,~\tau_0,~s$).  A great fraction of 
the surface of a caustic ring ($0 < |{\tau_1 \over \tau_0}| < {3 \over 4}$)
is saddle-shaped and therefore has tangent directions along which the 
curvature of the surface vanishes.  Finally, in section II, we 
derived the mass density profile near a cusp caustic.

Gravitational lensing was discussed in section III.  In general, 
the shift in image position is the gradient of a potential 
whose 2-dim Laplacian is the column density.  For an arbitrary 
mass distribution, the procedure for calculating the shift 
involves two steps.  First the matter density is integrated 
along the line of sight to obtain the matter density.  Second, 
the potential is obtained by convoluting the column density 
with the 2-dim. Green's function.  In the case of lensing by 
dark matter caustics, this procedure can be simplified: the 
shift is expressed directly as an integral over the parameter
space of the dark matter flow forming the caustic.  The relevant 
result is given in Eqs. (\ref{I}) and (\ref{LCI1}) if the 
caustic has contrast in the two dimensions transverse to the 
line of sight, and in Eq. (\ref{theeq}) if the caustic has contrast 
in only one of the dimensions transverse to the line of sight. 

In section IV, we applied our formalism to the gravitational lensing 
by dark matter caustics in four specific cases.  In the first three, 
the line of sight is tangent to a surface where a simple fold 
catastrophe is located.  The three cases are distinguished by the
curvature of the caustic surface at the tangent point in the direction 
of the line of sight: 1) the surface curves toward the side with two extra
flows, 2) the surface curves away from the side with two extra flows,
and 3) the surface has zero curvature.  In the fourth case (4)
studied, the line of sight is at a specific location close to a
cusp catastrophe and parallel to the plane of the cusp.  We found 
that each case has characteristic lensing signatures.  In three of 
the cases (2, 3 and 4) there are multiple images and infinite
magnification of point sources when their images merge.  In case 1, 
there are no multiple images.  Unfortunately, the effects are small 
even for dark matter caustic lenses at cosmological distances.  The
multiple images of a point source cannot be resolved with present 
instruments.  Typical magnifications and image distortions are  
of order one \% to a few \%.  
 
\section{Acknowledgments}

We would like to thank Stuart D. Wick and Achim Kempf for stimulating
discussions. This work was supported in part by the U.S. Department of 
Energy under grant DE-FG02-97ER41209, and by the National Science
Foundation under grant PHY99-07949.


\begin{figure}
\vspace{3.cm} 
\epsfxsize=5in
\centerline{\epsfbox{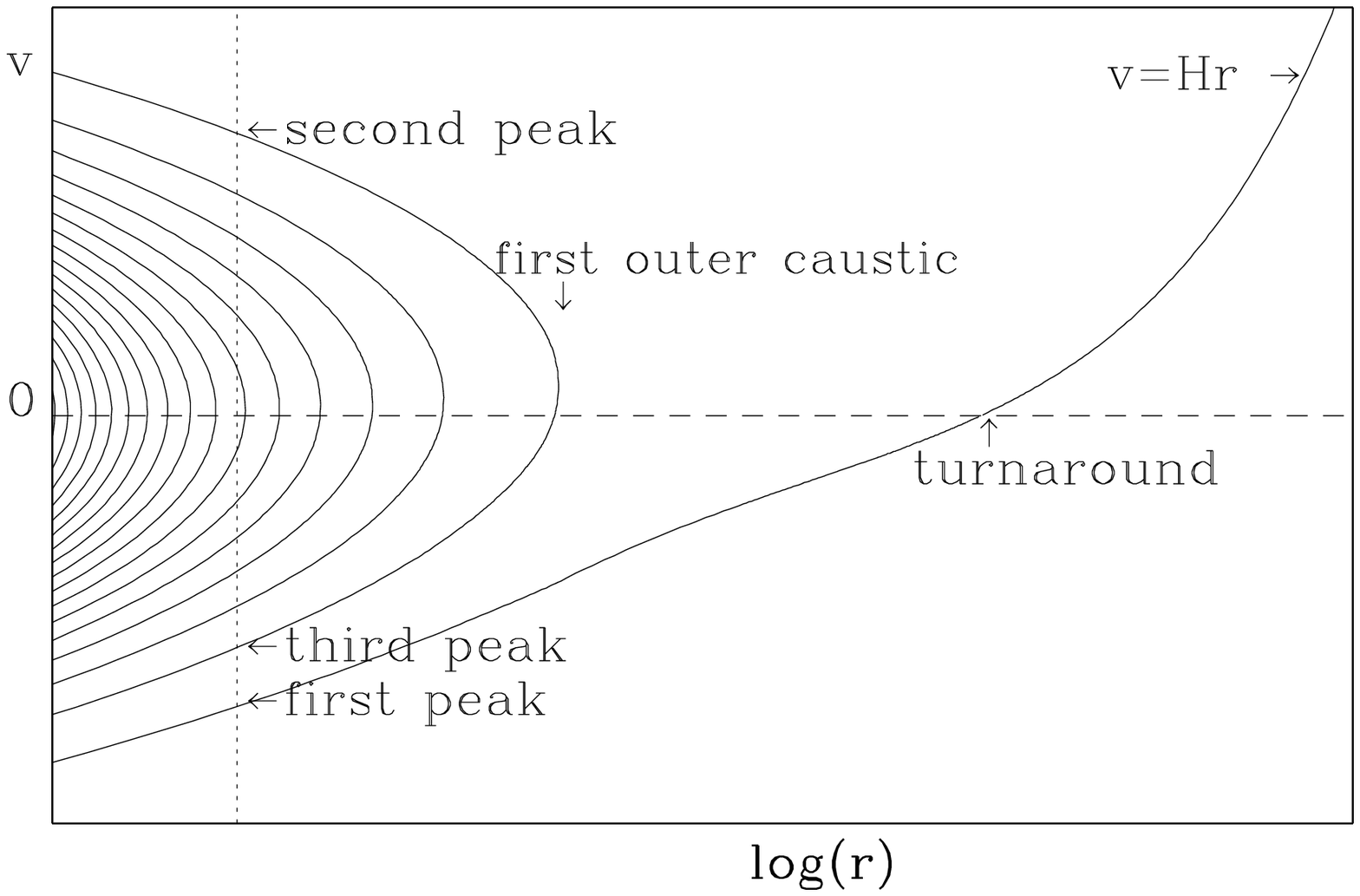}}
\vspace{1.cm}
\caption{A snapshot of the phase-space distribution of CDM particles
in a galactic halo.  For simplicity, spherical symmetry is assumed. r
is galactocentric distance and v is radial velocity.
The solid line indicates the location of the particles.  The dotted 
line corresponds 
to the observer position. Each intersection of the solid and dotted lines
corresponds to a CDM flow at the observer's location.  ``Turnaround''
refers to the moments in a particle's history when it has zero radial
velocity with respect to the galactic center. A ``caustic'' appears
wherever the 
phase-space line folds back. Particles pile up and hence the density
diverges there. }
\label{fig:phase} 
\end{figure} 

\begin{figure} \vspace{1.cm} \epsfxsize=6in
\centerline{\epsfbox{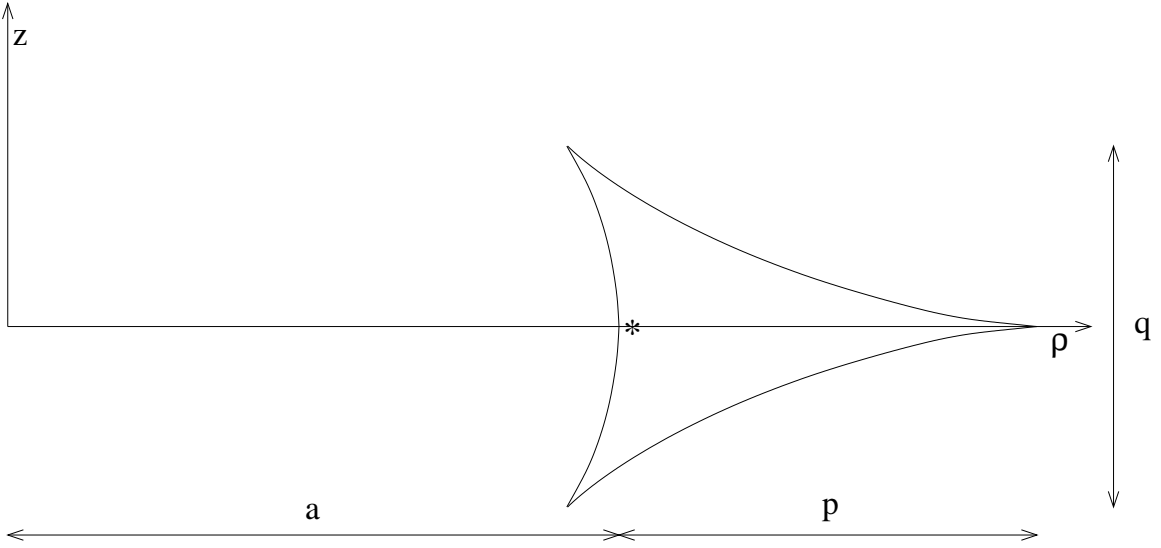}}
\vspace{1.cm}
\caption{Cross-section of a caustic ring in the case of axial and
reflection symmetry. p and q are the transverse dimensions in the 
$\hat{\rm \rho}$ and $\hat{\rm z}$ directions respectively. a is 
the ring radius. The star indicates the sample location ($\rho\simeq a ,
z=0$) discussed in section II.B.1. For clarity, we took $p,q \sim a$. 
For actual caustic rings, $p,q << a$.} 
\label{fig:tricusp} 
\end{figure} 

\begin{figure}
\vspace{1.cm} 
\epsfxsize=6in
\centerline{\epsfbox{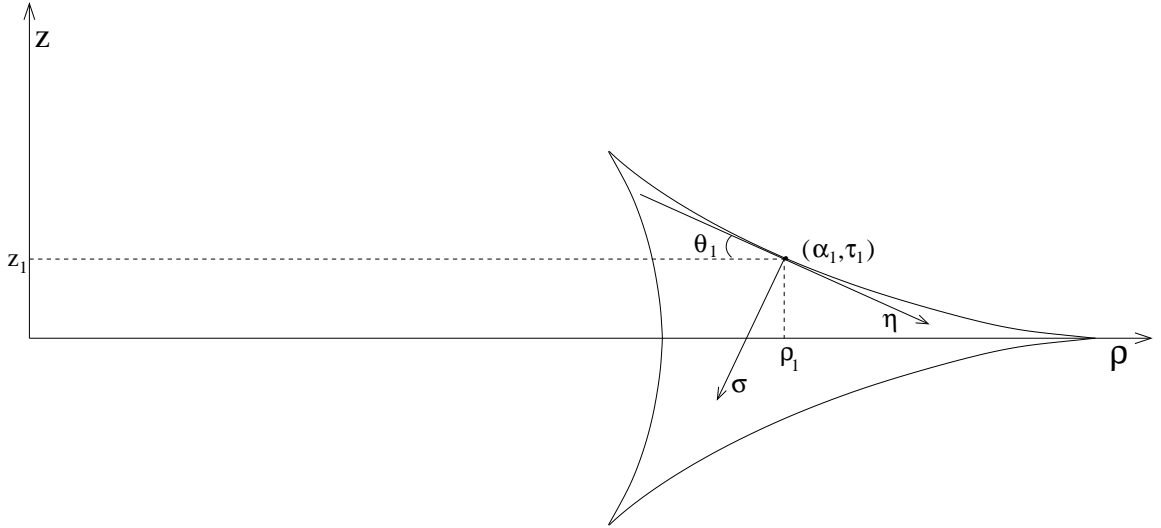}}
\vspace{1.cm}
\caption{An arbitrary point on the tricusp is labeled by $\tau_1$ 
with $\alpha_1 = \alpha_1 (\tau_1)$.  Its physical coordinates are
$(\rho_1, z_1)$. A new Cartesian coordinate system ($\sigma , \eta$) is
defined there such that $\hat{\sigma}$ is perpendicular to 
the caustic surface.  It is rotated relative to the $(\rho, z)$ 
cooordinates by an angle $\theta(\tau_1)+{{\pi}\over 2}$.}
\label{fig:tricusp1}
\end{figure}

\begin{figure}
\vspace{3.cm} 
\epsfxsize=5in
\centerline{\epsfbox{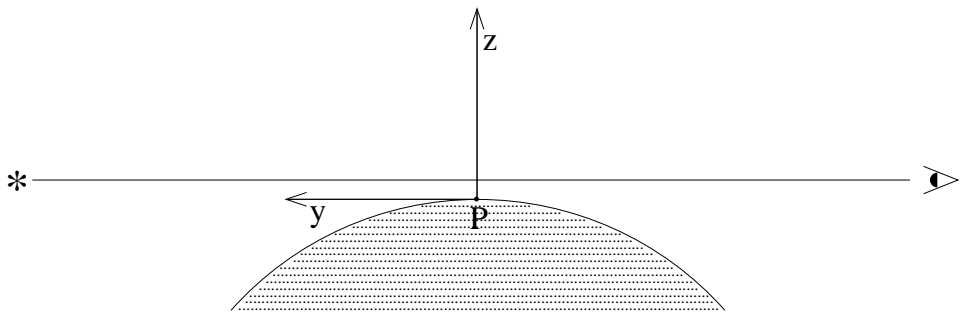}}
\vspace{1.cm}
\caption{Lensing by a concave fold. The arc is the intersection 
of the caustic surface with the plane containing the normal ($\hat{z}$)
to the surface and the line of sight ($\hat{y}$).  The shaded area 
indicates the side with the two extra flows.}
\label{fig:concave} 
\end{figure}

\begin{figure}
\vspace{1.cm} 
\epsfxsize=5in
\centerline{\epsfbox{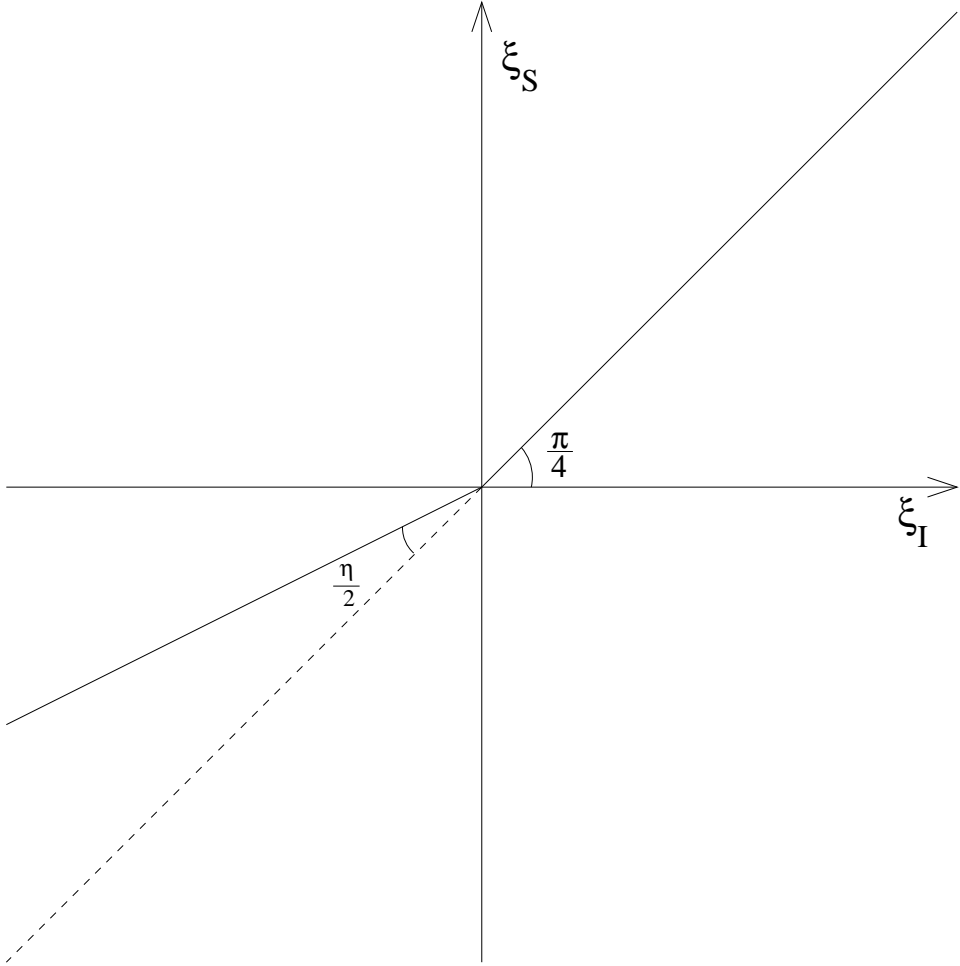}}
\vspace{1.cm}
\caption{Source position $\xi_S$ as a function of image position 
$\xi_I$ for lensing by a concave fold. $\eta$ is given 
by Eq. (\ref{ep}) in terms of the fold coefficient and the curvature
radius of the caustic surface.  Estimates of $\eta$ for the outer 
caustics of galactic halos are given in Eq. (\ref{epe}).}
\label{fig:shiftconcave}
\end{figure}

\begin{figure}
\vspace{1.cm}
\epsfxsize=5in
\centerline{\epsfbox{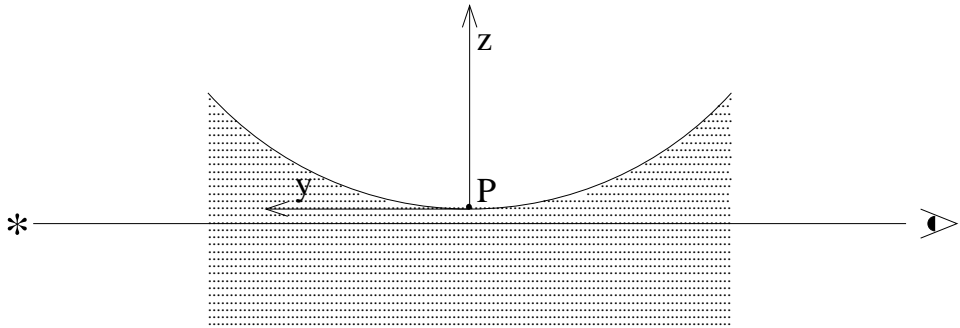}}
\vspace{1.cm}
\caption{Lensing by a convex fold.  Same as Fig. \ref{fig:concave}
except that now the caustic surface curves away from the side with two 
extra flows.}
\label{fig:convex}
\end{figure}

\begin{figure}
\vspace{1.cm} 
\epsfxsize=5in
\centerline{\epsfbox{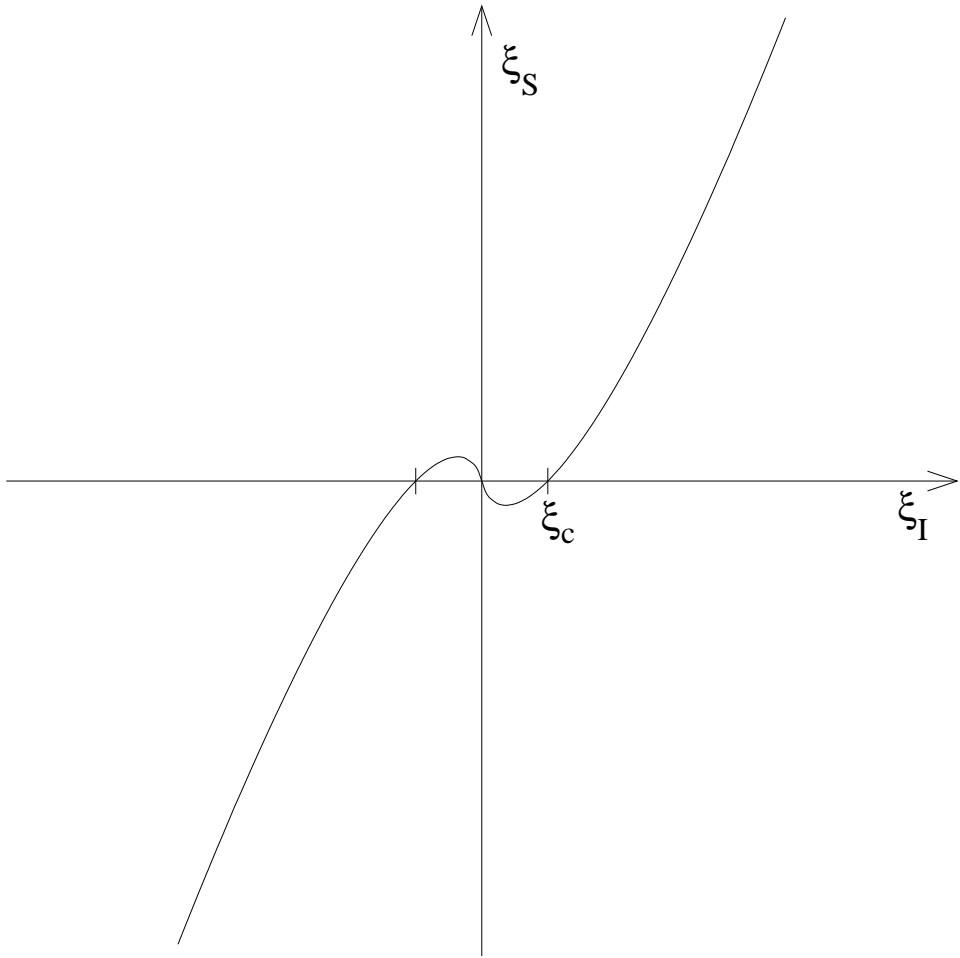}}
\vspace{1.cm}
\caption{Source position $\xi_S$ as a function of image position 
$\xi_I$ for lensing by a convex fold.  $\xi_c$ is given in
Eq. (\ref{xic}).}
\label{fig:shiftconvex}
\end{figure}

\begin{figure}
\vspace{1.cm} 
\epsfxsize=5in
\centerline{\epsfbox{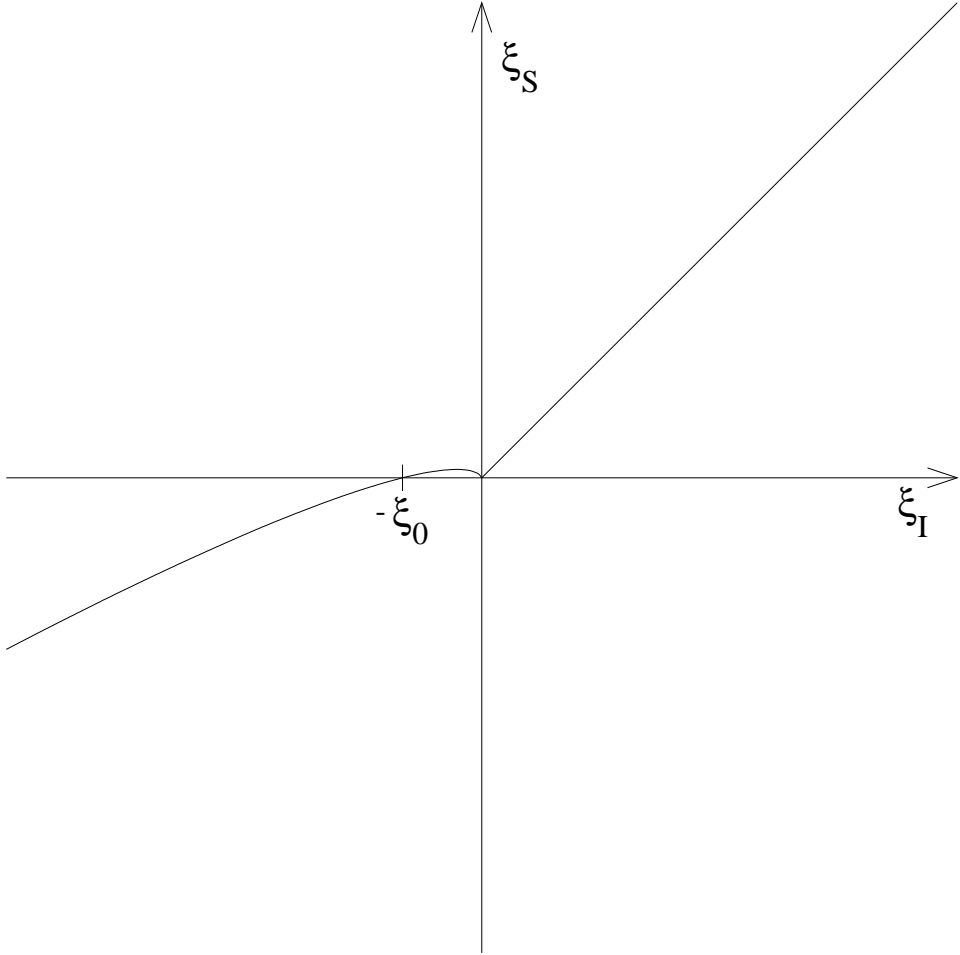}}
\vspace{1.cm}
\caption{Source position $\xi_S$ as a function of image position $\xi_I$
for lensing by a fold with zero curvature. $\xi_0$ is given 
in Eq. (\ref{xi0}).}
\label{fig:shiftflat} 
\end{figure}

\begin{figure}
\vspace{1.cm}
\epsfxsize=5in
\centerline{\epsfbox{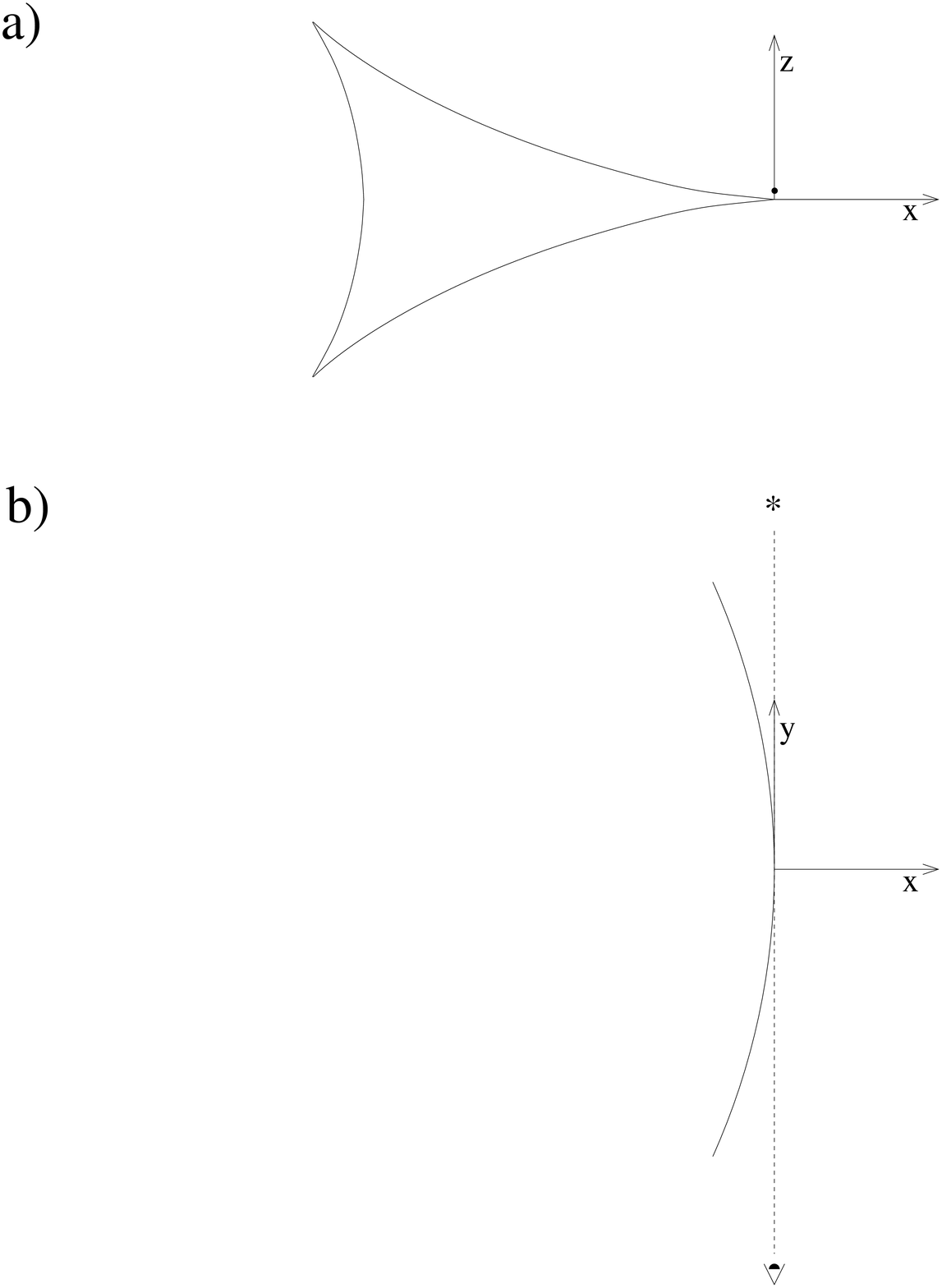}}
\vspace{1.cm}
\caption{Lensing by the caustic ring cusp at ($z, \rho) = (0, \rho_0$),
with line of sight parallel to the $z=0$ plane.  We define 
$x \equiv \rho - \rho_0$.  a) Side view in the direction of the line of
sight.  The latter is represented by the dot near $x=z=0$.  b)  Top
view.  The curve is the location of the cusp in the $z=0$ plane.} 
\label{fig:tricusp2} 
\end{figure} 

\begin{figure}
\vspace{1.cm}
\epsfxsize=6in
\centerline{\epsfbox{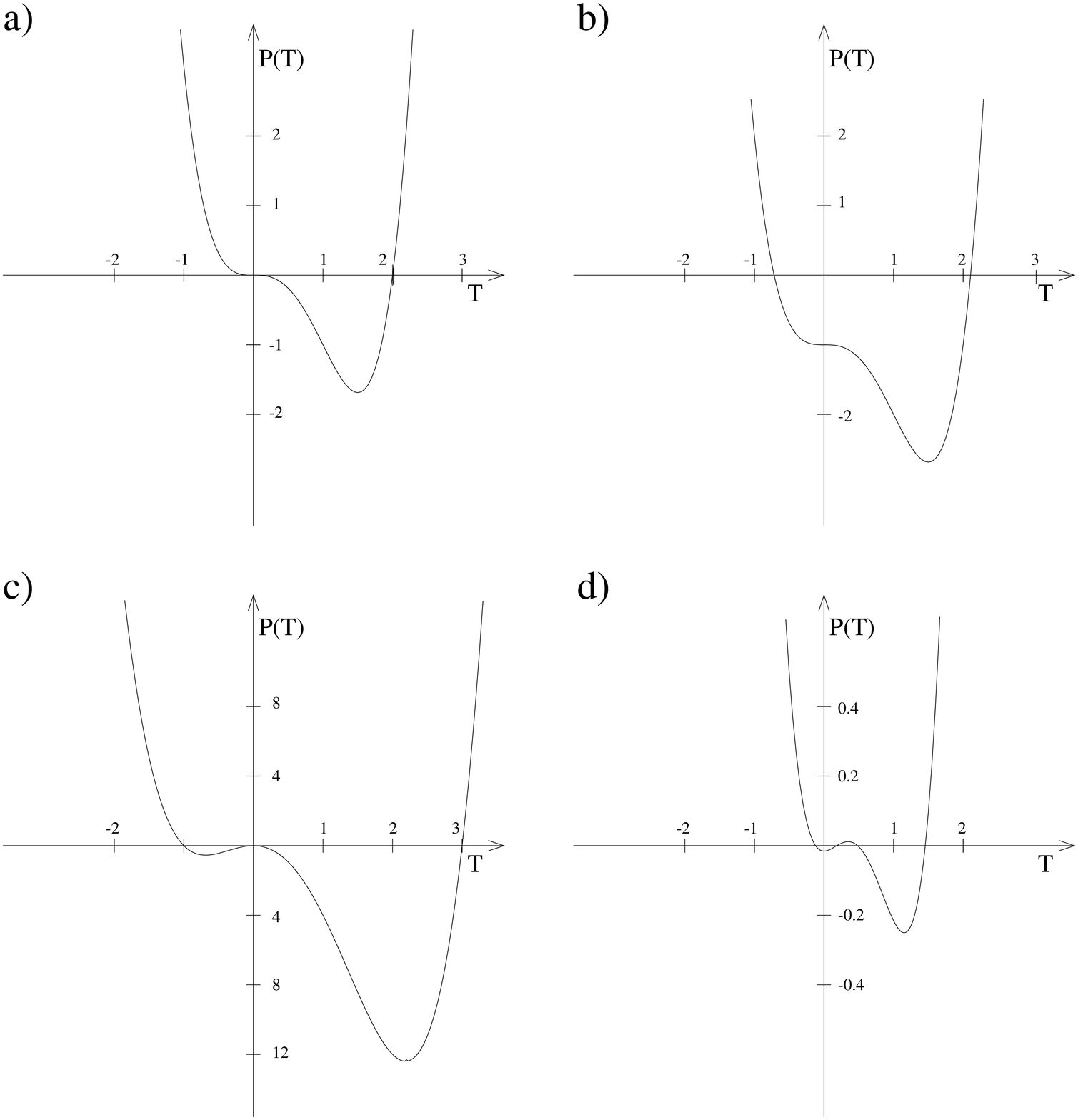}}
\vspace{1.cm}
\caption{ Graphs of the quartic polynomial 
$P(T)\equiv ~T^4-\zeta(2T^3+XT^2)-\frac{\zeta Z^2}{4}~$ for $\zeta =1$
and $(X , Z)$ =
$(0 , 0)$, $(0 , 2)$, $(1/3 , 0)$ and $(-1/3 , 0)$; a) through d)
 respectively.}
\label{fig:quartic} \end{figure}

\begin{figure}
\vspace{1.cm}
\epsfxsize=6in
\centerline{\epsfbox{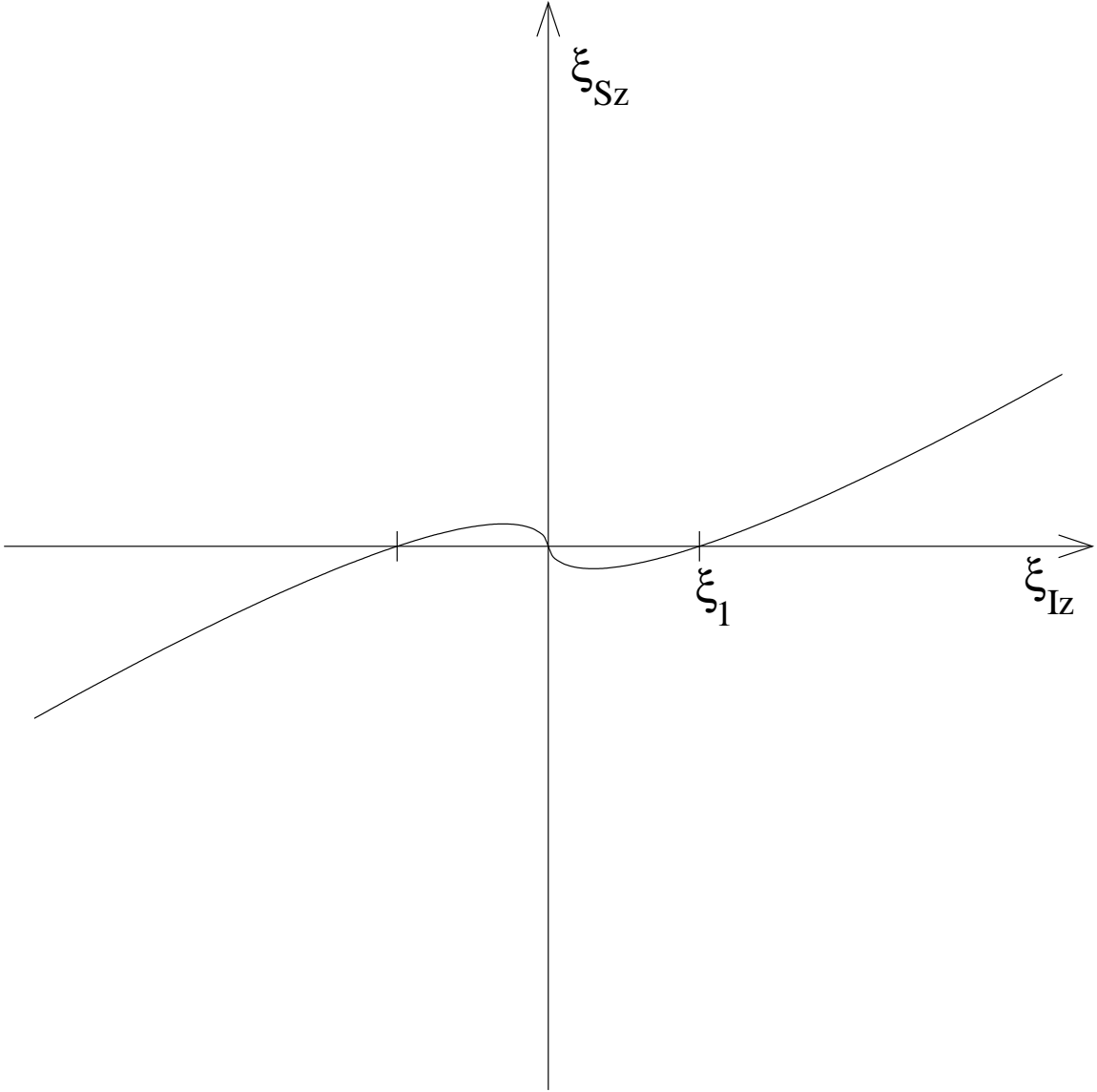}}
\vspace{1.cm}
\caption{Source position $\xi_{Sz}$ as a function image position 
$\xi_{Iz}$ for lensing by a cusp along the line of sight described 
in Fig. \ref{fig:tricusp2}.  $\xi_1 \sim \eta''^3$ where $\eta''$ 
is given by Eq. (\ref{ep''}).}
\label{fig:shiftcusp} \end{figure}

\end{document}